\documentclass[twocolumn,prb,aps,superscriptaddress,showpacs]{revtex4-1}   %

\usepackage{amsmath, amssymb}
\usepackage{bbm}
\allowdisplaybreaks
\usepackage{pdfpages}
\usepackage{color}
\usepackage{bm}
\usepackage{xr}
\usepackage{multirow}
\usepackage[caption=false]{subfig}
\usepackage{graphicx}
\usepackage{hyperref}
\usepackage[figure,table]{hypcap}               
\hypersetup{
pdftitle = {},
pdfsubject = {},
pdfauthor = {},
pdfkeywords = {},
pdfcreator = {},
pdfproducer = {LaTeX with hyperref},
colorlinks = true,
linkcolor = blue, 
anchorcolor = blue,
citecolor = blue, 
filecolor = red, 
menucolor = red, 
pagecolor = red, 
urlcolor  = blue, 
breaklinks = true,
pdfstartview = FitV,
pdfhighlight = /I,
pdfpagelayout = OneColumn,
hypertexnames=true 
}

\usepackage{color}                                      
\definecolor{d_red}{cmyk}{0.00, 0.81, 1.00, 0.27}
\definecolor{d_orange}{cmyk}{0.00, 0.33, 1.00, 0.00}
\definecolor{d_blue}{cmyk}{0.78, 0.47, 0.00, 0.20}
\definecolor{d_lgreen}{cmyk}{0.07, 0.00, 0.79, 0.29}
\definecolor{d_green}{cmyk}{0.66, 0.00, 0.71, 0.56}
\definecolor{d_blue}{cmyk}{0.78, 0.47, 0.00, 0.20}
\definecolor{d_dblue}{cmyk}{0.91, 0.79, 0.00, 0.22}
\definecolor{d_pink}{cmyk}{0.0, 0.79, 0.37, 0.29}
\definecolor{d_purple}{cmyk}{0.16, 0.54, 0.00, 0.70}
\definecolor{d_paleblue}{cmyk}{0.669, 0.338, 0.00, 0.373}
\definecolor{d_dpaleblue}{cmyk}{0.441, 0.290, 0.00, 0.580}
\definecolor{d_brown}{cmyk}{0.0, 0.490, 0.930, 0.350}
\definecolor{d_turquoise}{cmyk}{0.630, 0.04, 0.0, 0.440}

\newcommand{\av}[1]{\langle #1 \rangle}

\newcommand{\bfgg}{{\boldsymbol{G}}}

\newcommand{\bfqq}{{\boldsymbol{Q}}}

\newcommand{\bfss}{{\boldsymbol{S}}}

\newcommand{\bfa}{{\boldsymbol{a}}}
\newcommand{\bfb}{{\boldsymbol{b}}}

\newcommand{\bfn}{{\boldsymbol{n}}}

\newcommand{\bfr}{{\boldsymbol{r}}}

\newcommand{\bft}{{\boldsymbol{t}}}

\def\bmx{\begin{pmatrix}}
\def\emx{\end{pmatrix}}

\begin{document}
\title{Classical Heisenberg and planar spin models on the windmill lattice}
\author{Bhilahari Jeevanesan}
\affiliation{Institute for Theory of Condensed Matter, Karlsruhe Institute of Technology (KIT), 76131 Karlsruhe, Germany} 
\author{Peter P. Orth}
\affiliation{Institute for Theory of Condensed Matter, Karlsruhe Institute of Technology (KIT), 76131 Karlsruhe, Germany} 
\begin{abstract}
We investigate the classical Heisenberg and planar (XY) models on the windmill lattice. The windmill lattice is formed out of two widely occurring lattice geometries: a triangular lattice is coupled to its dual honeycomb lattice. Using a combination of iterative minimization, heat-bath Monte Carlo simulations and analytical calculations, we determine the complete ground state phase diagram of both models and find the exact energies of the phases. The phase diagram shows a rich phenomenology due to competing interactions and hosts, in addition to collinear and various coplanar phases, also intricate non-coplanar phases. We briefly outline different paths to an experimental realization of these spin models. Our extensive study provides a starting point for the investigation of quantum and thermal fluctuation effects. 
\end{abstract}
\maketitle

\section{Introduction}
\label{sec:introduction}
Insulating materials that host localized spin degrees of freedom can exhibit complex ground states and fascinating low-temperature properties. This behavior frequently arises from competing interactions that cannot be satisfied simultaneously. Prime examples are antiferromagnetic nearest-neighbor spin couplings on frustrated geometries like the two-dimensional (2D) triangular and kagome lattice or the three-dimensional (3D) pyrochlore lattice~\cite{Ramirez-FrustratedMagnets-AnnRevMatSci-1994, PhysRevLett.68.855, PhysRevLett.80.2929}. These systems are characterized by a large degeneracy of classical ground states. This often leads to complex states of matter and phase transitions if quantum or thermal fluctuations are present~\cite{PhysRevLett.62.2056,Shender82,PhysRevLett.110.077201}. 

A triangular lattice geometry with antiferromagnetic spin couplings is realized in a large number of magnetic materials such as $\text{Cs}_2 \text{Cu} \text{Cl}_2$~\cite{PhysRevLett.86.1335, PhysRevLett.88.137203, PhysRevLett.95.127202}, $\text{Na}_x \text{Co} \text{O}_2$~\cite{PhysRevB.56.R12685, PhysRevLett.92.247001}, $\text{Na} \text{Cr} \text{O}_2$~\cite{PhysRevLett.97.167203, PhysRevB.80.054406} and $\alpha-\text{Na} \text{Fe} \text{O}_2$~\cite{PhysRevB.76.024420}. Another frustrated triangular material is the recently discussed cluster magnet $\text{Li} \text{Zn}_2 \text{Mo}_3 \text{O}_8$~\cite{SheckeltonMcQueen-NatMat-2012, PhysRevB.89.064407, PhysRevLett.112.027202, PhysRevLett.111.217201}. Here, $\text{Mo}_3 \text{O}_{13}$ clusters that carry a total spin $S=1/2$ are arranged in two-dimensional triangular lattice planes that are weakly coupled along the third dimension. Another lattice geometry that exhibits frustration effects if further neighbor antiferromagnetic couplings are present is the honeycomb lattice. This lattice is dual to the triangular lattice. It is realized in various solid-state compounds and can arise by replacing one third of the magnetic ions in a triangular lattice system by a non-magnetic one. This is done, for example, in $\text{Na}_2 \text{Co}_2 \text{Te} \text{O}_6$ or $\text{Na}_3 \text{Co}_2 \text{Sb} \text{O}_6$~\cite{Viciu20071060}. Another recently discussed interesting honeycomb material is $\text{Na}_{1-x} \text{Ni} \text{Sb} \text{O}_6$, where magnetic $\text{Ni}^{2+}$ and $\text{Ni}^{3+}$ form a honeycomb lattice with mixed spins $S=1/2$ and $S=1$. 

Here, we study the situation where these two lattice geometries are combined and consider spins on a honeycomb lattice that are antiferromagnetically coupled to spins situated on a triangular lattice. It has recently been shown that such a setup shows intriguing order from disorder effects in the regime of weakly coupled sublattices. There, an emergent $\mathbb{Z}_6$ degree of freedom has been revealed that exhibits a sequence of Berezinskii-Kosterlitz-Thouless phase transitions bracketing a critical phase~\cite{PhysRevLett.109.237205,PhysRevB.89.094417}. Experimentally, such a situation might arise, for example, in the cluster magnet $\text{Li} \text{Zn}_2 \text{Mo}_3 \text{O}_8$ at low temperatures~\cite{PhysRevLett.111.217201} or in a material such as $\text{Na}_{1-x} \text{Ni} \text{Sb} \text{O}_6$ by replacing the non-magnetic ion $\text{Sb}$ by a magnetic one.  The progress in chemical synthesis, in particular considering the approach of using small magnetic clusters as basic units, might bring other experimental candidates in the future as well, possibly with a large spin $S > 1/2$. 
\begin{figure}[b]
\includegraphics[width=\linewidth]{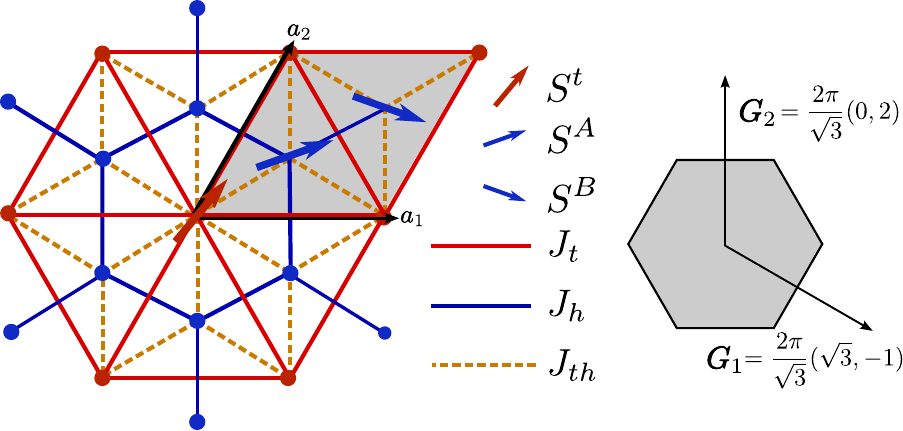}
\caption{Windmill lattice. Left: The lattice may be thought of as a triangular lattice (red vertices) together with its dual honeycomb lattice (blue vertices). The model has a three-site basis. The Bravais lattice vectors are ${\bm a}_1$ and ${\bm a}_2$. The shaded region is a unit cell. Right: Reciprocal lattice vectors with first Brillouin zone.}
\label{model}
\end{figure}

Another experimental platform where classical frustrated magnetism has been investigated in recent years is based on cold bosonic atoms in optical lattices~\cite{Struck19082011, StruckSengstockMathey-NatPhys-2013}. At low temperatures and for weak interactions these systems form a superfluid and the atoms at site $i$ in the lattice have a well-defined local phase $\phi_i$. This local phase degree of freedom can be interpreted as a classical planar (XY) spin $\bigl( \cos \phi_i, \sin \phi_i \bigr)$. Nearest-neighbor spins are coupled via tunneling of atoms between the sites. For regular tunneling of atoms between the sites, the associated coupling between the planar spins is ferromagnetic. By shaking the lattice in a periodic way, however, it is possible to add a non-zero Peierls phase to the tunneling element. In this way, it is possible to induce a change in the sign of the tunneling element which leads to an antiferromagnetic coupling between the spins. Within this approach, different links of the lattice can be addressed independently. Frustration effects have been experimentally observed on the triangular lattice via standard time-of-flight imaging~\cite{Struck19082011}. Honeycomb optical lattice geometries have also been realized in the past~\cite{SoltanPanahiSengstock-NaturePhys-2011, EsslingerDirac-Nature2012}. 

This serves as our motivation to extensively study the classical Heisenberg and the classical planar (XY) spin model on a lattice that combines both a honeycomb and a triangular lattice. We refer to this lattice, which is shown in Fig.~\ref{model}, as the ``windmill lattice''. Considering antiferromagnetic interactions between all nearest-neighbor pairs of spins, we determine the complete ground state phase diagram of both models. Due to competing interactions, the models turn out to show an extremely rich ground state phenomenology. For the Heisenberg model we find that next to phases where the spins order in a collinear or a coplanar fashion there exist also phases where the spins exhibit an intricate non-coplanar configuration where they arrange themselves into seperate double cones. In the windmill XY model those non-coplanar phases are replaced by other similarly involved configurations.

The structure of the remainder of the paper is as follows: in Sec.~\ref{sec:model} we introduce the classical Heisenberg and planar spin models on the windmill lattice, and in Sec.~\ref{sec:method} we describe the methods that we employ to obtain the ground state phase diagram. We use an ``iterative minimization'' technique to find a variational expression of the ground state whose energy can be analytically computed, minimized and compared to the numerical result. In Sec.~\ref{sec:zero-temp-phase} we present one of our main results: the full ground state phase diagram of the Heisenberg model on the windmill lattice as a function of exchange couplings. In Sec.~\ref{sec:phases}, we then discuss the various ground state phases in detail. In Sec.~\ref{sec:xy-model} we analyze the planar (XY) model on the windmill lattice and determine its complete ground state phase diagram. In the appendices we provide the details on the calculation of all the ground state energies from the variational forms.

\section{Windmill spin model}
\label{sec:model}
The windmill lattice that we study consists of a triangular lattice combined with its dual, honeycomb lattice. The windmill lattice can be described as a triangular Bravais lattice with a three-site basis per unit cell containing triangular, honeycomb $A$ and $B$ sites. The spins are positioned on the vertices of the lattice as shown in Fig.~\ref{model}. We consider classical spins with an antiferromagnetic exchange coupling between all nearest-neighbor pairs of spins. 
The Hamiltonian of the windmill model reads
\begin{eqnarray}
H&=&J_t \sum_{\langle ij \rangle} {\bm S}^{t}_i \cdot {\bm S}^{t}_j + J_{{h}} \sum_{\langle ij \rangle} {\bm S}^{A}_i \cdot {\bm S}^{B}_j \nonumber\\
&&+ J_{{th}} \sum_{\langle ij \rangle,\nu=A,B} {\bm S}^{t}_i \cdot {\bm S}^{\nu}_j \label{hamiltonian}
\end{eqnarray}
where $\av{i,j}$ denotes summation over each nearest-neighbor pair, the index $\{t, A, B\}$ refers to the sublattices and $J_t,J_{{h}},J_{{th}}$ are positive, \emph{i.e.}, antiferromagnetic, coupling constants. In the classical windmill model all spins are classical unit vectors. The vectors have three components in the case of Heisenberg spins $\bfss^\alpha_i = (S^\alpha_{i,x}, S^\alpha_{i,y}, S^\alpha_{i,z})$ and two components in the case of planar (XY) spins $\bfss^\alpha_i = (S^\alpha_{i,x}, S^\alpha_{i,y})$ with $\alpha \in \{t, A, B\}$.

The triangular Bravais lattice with lattice constant $a$ is spanned by the primitive lattice vectors $\bfa_1 = a(1, 0)$ and $\bfa_2 = \frac{a}{2} (1, \sqrt{3})$.
The basis vectors are given by $\bfb_t =(0,0)$, $\bfb_A = \frac{1}{3} \bfa_1+\frac{1}{3} \bfa_2$ and $\bfb_B = \frac{2}{3} \bfa_1+\frac{2}{3} \bfa_2$.  
The reciprocal vectors take the form $\bfgg_1 = \frac{2 \pi}{\sqrt{3} a} (\sqrt{3}, - 1)$ and $\bfgg_2 = \frac{2 \pi}{\sqrt{3} a} ( 0, 2)$, and are shown in Fig.~\ref{model}.

\section{Methodology}
\label{sec:method}
All the results in this paper were obtained by using an ``iterative minimization'' algorithm that has been employed in the literature to discover ground state configurations of classical spin models~\cite{PhysRevB.88.024407,LapaHenley-arXiv-2012}. Independently we verified our results by using heat-bath Monte Carlo simulations~\cite{Miyatake} in combination with parallel tempering updates~\cite{MarinariParisi,HukushimaNemoto}. We then extract variational forms of the spin configurations and determine the variational parameters by minimizing the corresponding ground state energies. We analytically find the configuration of minimal energy which determines the phase diagram and the phase boundaries. 

We begin with an explanation of the iterative minimization algorithm. Starting from a randomized spin configuration, in every iteration of the algorithm a spin is chosen at random and rotated such as to minimize the interaction energy with its neighbors. Each step of the algorithm is an update of the form
\begin{eqnarray}
\bm{S}_{i}^{t}\rightarrow\bm{S}_{i}^{t}=-\frac{J_{t}\sum_{k}\bm{S}_{k}^{t}+J_{{th}}\sum_{k,\nu}\bm{S}_{k}^{\nu}}{\Vert J_{t}\sum_{k}\bm{S}_{k}^{t}+J_{{th}}\sum_{k,\nu}\bm{S}_{k}^{\nu}\Vert} \\
\bm{S}_{j}^{\alpha}\rightarrow\bm{S}_{j}^{\alpha}=-\frac{J_{h}\sum_{k,\nu}\bm{S}_{k}^{\nu}+J_{{th}}\sum_{k}\bm{S}_{k}^{t}}{\Vert J_{h}\sum_{k,\nu}\bm{S}_{k}^{\nu}+J_{{th}}\sum_{k}\bm{S}_k^{t}\Vert}
\end{eqnarray}
with the index $\alpha \in\{A,B\}$. The sum over index $\nu = A,B$ runs over both honeycomb sublattices and the summation over the index $k$ ranges over the neighbors of the spin that is being updated. This technique does not provide rigorous proofs for the correctness of the discovered phases. One difficulty that one may imagine is that the algorithm converges to a local minimum of the energy landscape. In this case a local update is not capable of improving the energy. This possibility is made unlikely by the fact that for all phases we ran the algorithm multiple times with different random initial configurations and observed that the system always converged to the same phase.

In order to discover the groundstate configurations, the algorithm was applied to a spin system with ${N=L\times L=30\times 30}$ unit cells (\emph{i.e.}, $N$ triangular spins and $2N$ honeycomb spins) with periodic boundary conditions. We applied the minimization algorithm for 50,000 steps. The resulting spin configurations were all explored and a mathematical description of their spin ordering was extracted, from which the energies were analytically calculated.

After the numerical search, we performed systematic runs on a lattice with $N=12\times12$ unit cells to confirm the absence of further phases. This run was performed for the regime of parameters ${J_{{h}} / J_t \in [0.2,\dots,9.4]}$ in steps of $0.1$ and  ${J_{{th}} / J_t\in [0.2,\dots,9.4]}$ in steps of $0.1$. At every point within this parameter range, the minimization algorithm was applied for $60,000\times N$ iterations. Every application optimized one randomly chosen triangular lattice spin and one randomly chosen honeycomb lattice spin. The resulting energies of the converged configurations were compared to the analytically computed energies. 

Some of the discovered phases have spin configurations that depend parametrically on the coupling constants such as the pitch angle in a spiral phase. Here, we computed the energy by leaving these variables, like the pitch angle, as variational parameters and obtained their value by minimizing the energy with respect to the parameters. 
In most cases, the results of the simulation were highly converged such that the energy per spin that emerged from the simulation numerically coincided exactly with the energy computed by minimization of the variational state. Similarly, the spin configurations (e.g. scalar products between neighboring spins) of all the simulated phases were to many digits identical to those of the phases proposed.  In the few cases where the convergence was not so good, which was evidenced by the higher energy per spin value, we investigated snapshots of the spin configurations and found that the algorithm had been trapped in a local minimum with topological defects that could not be removed by local updates. Neverthless, even in the presence of these defects the snapshots still showed ordering of the spins that were clearly those of the proposed phases.

As a final test we repeated the systematic run with Monte Carlo simulations. The simulations were carried out at $40$ temperature points $T_i$ covering the interval $[10^{-3}J_t, \dots, 2.0 J_t]$. The temperature points are chosen to be geometrically spaced~\cite{Katzgraber}, \emph{i.e.}, $T_{i+1}/T_i$ is a constant ratio. The lattice has ${N=L\times L=30\times 30}$ unit cells and we again employ periodic boundary conditions. The simulations are done in parallel for all temperatures with the spin configurations of the $40$ lattices stored simultaneously. To every one of these lattices the heat-bath algorithm is applied $N$ times. This is followed by a parallel tempering move. The latter kind of update consists in proposing for every pair of neighboring temperatures $(T_i,T_{i+1})$ an exchange of the full spin configurations. The proposals are accepted/rejected according to the standard Metropolis-Hastings rule. The parallel tempering algorithm helps to quickly produce uncorrelated spin configurations. 

The cycle of heat-bath steps followed by parallel tempering updates was repeated a total of $10,000$ times before we finally measured the energy of the lattice with the smallest temperature. 

In this way the energies were determined for the coupling constants in the parameter regime $j_{h}\equiv J_h/J_t\in [0.1,\dots, 10.0]$ and $\bar{J} \equiv J_{th}/\sqrt{J_t J_h}\in[0.1, \dots, 4.0]$. We found all resulting energies to be slightly larger than the analytically calculated minimum energies. This is further confirmation of the absence of ground state phases other than the ones we have found. 

\section{Ground state phase diagram of Heisenberg windmill model}
\label{sec:zero-temp-phase}
\begin{figure}[h!]
\includegraphics[width=\linewidth]{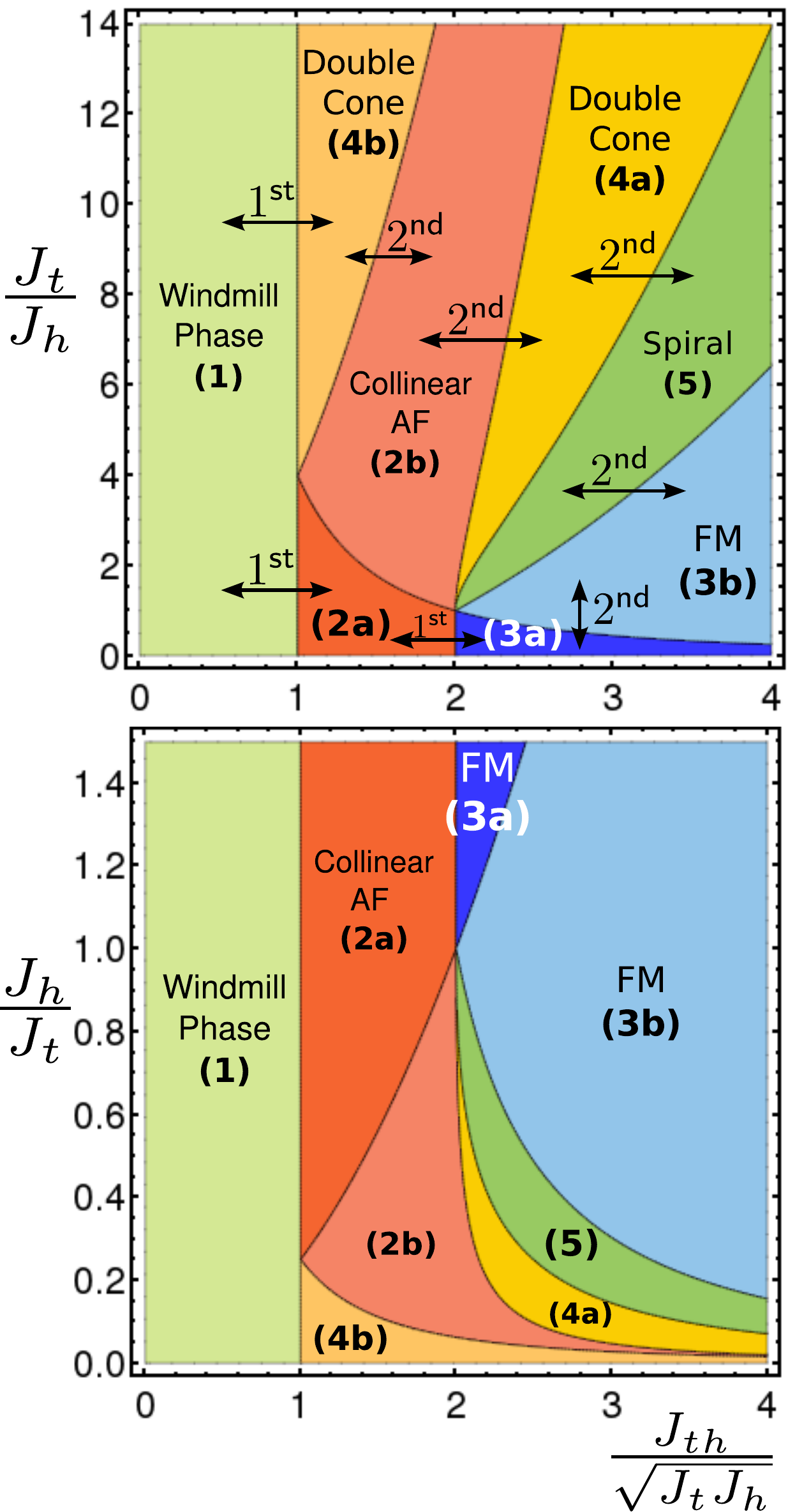}
\caption{Ground state phase diagram of the Heisenberg model on the windmill lattice as a function of the dimensionless ratios of coupling constants $J_t/J_h$ and $J_{th}/\sqrt{J_t J_h}$. In the upper figure we use the $y$-axis label $J_{t}/J_h$ while in the lower figure we use the inverse ratio $J_h/J_t$. The different phases are described in detail in Sec.~\ref{sec:phases}. The order of the phase transition between the phases is indicated above the arrows. The non-coplanar phases are labelled $(4a)$ and $(4b)$. }
\label{fig:2}
\end{figure}

The complete ground state phase diagram of the Heisenberg windmill model is shown in Fig.~\ref{fig:2} as a function of the ratios of coupling constants $J_t/J_h$ and $J_{th}/\sqrt{J_t J_h}$. Note that the vertical axis is the ratio $J_t/J_h$ in the upper panel of the figure, while we use the inverse ratio $J_h/J_t$ in the lower panel. The horizontal axis is $J_{th}/\sqrt{J_t J_h}$. In total, the Heisenberg model exhibits eight different ground state phases, which we describe in detail in Sec.~\ref{sec:phases}. 

The energies of the different spin configurations, measured in units of $J_t$, are functions of the dimensionless coupling constant ratios 
\begin{eqnarray}
 j_h &=& J_h/J_t    \label{eq:3} \\
 j_{th} &=& J_{th}/J_t. \label{eq:4}
\end{eqnarray}
The ground state energy $E(j_h, j_{th})$ is defined as the energy of the spin configuration with the lowest energy, and the ground state phase is described by this spin configuration.

As a function of the couplings $j_h$ and $j_{th}$ there will eventually be level crossings where the energies of the spin configurations with the lowest two energies switch places~\cite{goldenfeld_pt}. A phase transition occurs at a crossing of the energies of two different spin configurations. Exactly at the transition point, their energies match, but their (higher order) derivatives will generally not be the same. According to the usual Ehrenfest classification of phase transitions, the order of the phase transition is determined by the lowest order of the derivative of the ground state energy $E(j_{h}, j_{th})$ with respect to the tuning parameter, $j_h$ or $j_{th}$, that exhibits a singularity. In the phase diagram we have indicated the order of the phase transition by the labels $1^{\text{st}}$ and $2^{\text{nd}}$ for first and second order phase transitions. 

Some of the discovered phases are continuously connected to each other, and exactly at the level crossing the spin configurations are identical. We often denote those phases in the phase diagram by the suffix a or b. In this case both the energies as well as the first order derivatives of the energies will match at the transition point. The second order derivatives, however, will in general not be equal. The phase transitions between phases that are continuously deformed into each other at the transition are therefore always of second order. 

There are two special points in the phase diagram, where multiple phases become energetically degenerate. These are the points $(j_h, j_{th}/\sqrt{j_h}) = (1/4, 1)$ and at $(j_h, j_{th}/\sqrt{j_h}) = (1, 2)$. One would not expect a large number of unrelated phases to coincide at one point. In fact, one can prove a simple result similar to the Gibbs phase rule~\cite{Fermi} about the number of generically coinciding, energically degenerate phases. Let $E_1(j_h,j_{th}), E_2(j_h,j_{th})$ and $E_3(j_h,j_{th})$ be the energies of three minimum energy phases. The requirement that these energies should coincide at a point $(j_h,j_{th})$ is expressed by the conditions 
\begin{eqnarray}
E_1(j_h,j_{th})&=&E_2(j_h,j_{th})  \label{eq:1}\\
E_2(j_h,j_{th})&=&E_3(j_h,j_{th}) \label{eq:2} 
\end{eqnarray}
which can be expected to be a solvable system in $j_h$ and $j_{th}$, since the number of variables equals the number of conditions. Equating a larger number of energy functions will generally result in an overdetermined, unsolvable system, unless the energy functions are related to each other in a special way. Thus the number of generically coincident points is limited to three.

In the ground state phase diagram in Fig.~\ref{fig:2} there are, however, seven phases coinciding at the point $(j_h,j_{th}/\sqrt{j_h})=(1,2)$ and four phases coinciding at the point $(j_h,j_{th}/\sqrt{j_h})=(1/4,1)$. 
This is possible because a number of these phases are related to each other in the way described above, \emph{i.e.}, they continuously transform into each other at the degeneracy point. 

Let us consider the point $(j_h,j_{th}/\sqrt{j_h})=(1,2)$ in more detail. The detailed description of the different phases can be found below in Sec.~\ref{sec:phases}. Exactly at the point $(j_h,j_{th}/\sqrt{j_h})=(1,2)$ there are two energetically equal ground state configurations, which are those of phase $(2b)$ and phase $(3b)$ (see Secs.~\ref{sec:coll-antif-ferr} and~\ref{sec:ferromagnet-3}). In the limit $(j_h, j_{th}/\sqrt{j_h}) \rightarrow (1,2)$ the spin configurations of all the surrounding phases are equal to the configuration of one of these two phases. All these phases must, therefore, meet at this point in the phase diagram. Obviously phase $(2a)$ deforms into $(2b)$ and phase $(3a)$ into $(3b)$. The half-opening angles of the conical phase $(4a)$ (see Sec.~\ref{sec:double-cone-phase}) go to zero in this limit, which yields a spin configuration identical to the one of phase $(2b)$. Similarly, the spiral angle of phase $(5)$ (see Sec.~\ref{sec:incomm-spir-5}) goes to zero resulting in a spin configuration that is identical to the one of phase $(3b)$. 

A similar line of reasoning explains the degeneracy at the point $(j_h, j_{th}/\sqrt{j_h}) = (1/4,1)$. Here, phases $(4b)$ and $(2a)$ turn into $(2b)$. The opening angle $\alpha_t$ of the double cone configuration tends to zero as the border to region (2b) is approached. This border includes the degeneracy point, as shown in the phase diagram.

\section{Ground state phases}
\label{sec:phases}
In the following we describe the phases that were found in a broad search of the two-dimensional parameter space ${(j_h = J_{{h}} /J_t, j_{th}/\sqrt{j_h} = J_{{th}}/\sqrt{J_t J_{{h}}})}$. The phases are labelled as shown in the phase diagram in Fig.~\ref{fig:2}. In describing these phases we adopt the convention of placing the spins in the coplanar phases in the $S_x$-$S_y$-plane in spin space.  In many of the following phases there is an additional degeneracy in that a certain symmetry may be broken along different directions of lattice space. In such cases we adopt one particular direction for the description of the phase and follow this with a discussion about the symmetry properties. Next to the figures with the spin arrangements are shown the positions of the ordering wave vectors in the Brillouin zone with the wave vectors of the honeycomb lattice as blue points and triangular lattice as red points. 
\begin{table}[b!]
\centering
\begin{tabular}{c|c}
\hline\hline
{Phase} & Order Parameter Manifold (Heisenberg model)\\
\hline
1 & $SO(3)\times O(3)/O(2)$ \\
2a, 2b & $O(3)/O(2)\times O(2) \times \mathbb{Z}_3$\\
3a & $O(3)/O(2)\times O(2)$ \\ 
3b &  $O(3)/O(2)$ \\ 
4a &  $SO(3)\times \mathbb{Z}_3$ \\  
4b &  $SO(3)\times \mathbb{Z}_3$ \\ 
5 & $ SO(3)\times {\mathbb Z}_3$   \\
\hline\hline
\end{tabular}
\caption{Order parameter manifold of the ground state phases of the Heisenberg windmill model. The elements of these group manifolds transform energetically degenerate ground state spin configurations into each other.}
\label{tablesymm}
\end{table}

In some of the following ground state phases the symmetry of the lattice is broken. This can manifest itself, for example, in stripes of equal spin orientation along a certain direction in lattice space. In such cases, one can obtain a distinct, but energetically degenerate ground state by means of a rotation of all spins in lattice space around an arbitrary triangular site. The description of such a rotated state is complicated by the fact that the windmill lattice has a three site basis. The operation of rotating the lattice by $60^\circ$ will in general also involve a reattribution of honeycomb spins to their respective unit cells and a possible relabeling of A into B site spins and vice versa. In order to avoid such complications, our description in the following will always be of only one lattice configuration. The other symmetry related configurations are discussed. 

The differerent phases are characterized by their order parameter manifold. This manifold is defined as the symmetry group, whose elements transform a given spin configuration into a distinct, but energetically degenerate configuration. A simple (anti-)ferromagnetic state in the Heisenberg model thus has the order parameter manifold $O(3)/O(2)$, since it is characterized by a normalized unit vector in the direction of magnetization $\bfn$. Global rotations around an axis that is not (anti-)parallel to a given magnetization $\bfn$ yield another energetically degenerate ground state spin configuration. In Table~\ref{tablesymm} we list the order parameters manifolds of the various ground state phases. 

\begin{table}[b!]
\centering
\begin{tabular}{c|c|c}
\hline\hline
{Phase} & Energies $E/{NJ_t}$ & Condition  \\
\hline
1 & $-\frac{3}{2} -3 j_h$ & none\\
2a & $- 1 -3  j_h - \frac{j^2_{th}}{2j_h}$ & $2j_h\geq j_{th}$ \\
2b & $- 1 -  j_h-2j_{th}$ & $2j_h \leq j_{th}$ \\ 
3a & $3  - 3 j_h-\frac{3}{2} \frac{j^2_{th}}{J_h}$ & $2j_h \geq j_{th}$ \\ 
3b & $3  + 3j_h-6 j_{th}$ & $2j_h \leq j_{th}$ \\ 
4a & see  (\ref{energycone}) & $j_{th}^2\geq 4j_h$, $\sigma\leq1\leq\rho$ \\ 
4b & $-\frac{3}{2} -j_h -2 j_{th}^2$ & $2j_{th}\leq 1 $\\
5 & $ -\frac{3}{2}- \frac{(2j_{th}-j_h)^2} {2}$ &  $|1+j_h-2j_{th}|\leq 2$ \\
XY I & $-\frac{3}{2} -  \frac{(2j_{th} + j_h)^2 } {2}$ & $j_{th}+j_h /2 \leq 1/2 $ \\
XY II & see Eq.~\eqref{eq:35}) & $j_{th}^2\geq2j_h$, $\rho\leq1\leq\sigma$ \\
\hline\hline
\end{tabular}
\caption{Energies of the different ground state spin configurations of the Heisenberg and XY windmill model. }
\label{table_energies}
\end{table}

\begin{table}[b!]
\centering
\begin{tabular}{c|c|c}
\hline\hline
Phases & Phase Boundary & Order of transition  \\
\hline
1 : 2a & $j_{th}^2=j_h$ & $1$\\
1 : 4b &  $j_{th}^2=j_h$& $1 $  \\
2a : 2b &$j_{th}=2j_h$ & $2 $ \\ 
2b : 4b & $j_{th}=\frac{1}{2} $&  $2 $\\ 
2b : 4a &$ 2j_h=\sqrt{2}j_{th}^{3/2}-j_{th}$   & $2  $ \\ 
2a : 3a & $j_{th}^2 =4 j_h $ &  $1 $\\ 
3a : 3b &$ j_{th}=2j_h$  & $2$ \\
3b : 5 &$j_{th}=\frac{3}{2}+\frac{1}{2}j_h$ & $2$ \\
4a : 5 &$j_h+1=j_{th}  $ & $2$ \\
1: XY I &$6j_h= (2j_{th}+j_h)^2$ & $1$\\
1 : 2b & $2j_{th}=2j_h+1/2$ & $1$\\
XY I : 2b & $j_h+2j_{th}=1$ & 2\\
2b : XY II & $2j_{th}^2=j_{th}+j_{th}j_h+4j_h$ & 2\\
5 : XY II & $E_5=E_{\text{XY II}}$ & 1\\
\hline\hline
\end{tabular}
\caption{Parametric location of the phase boundaries and order of the phase transition between different phases in the ground state phase diagram of the Heisenberg and XY windmill model (see Figs.~\ref{fig:2} and~\ref{phasediag}).}
\label{boundaries}
\end{table}
We measure energies in units of $NJ_t$, where $N$ is the number of the spins on the triangular lattice. The energies of the ground state spin configurations can be analytically calculated and are given in Table~\ref{table_energies}. We refer to Appendix~\ref{sec:energies} for the details. Finally, in Table~\ref{boundaries} we list the functional forms of the boundaries between the phases, which are calculated from the explicit form of the energies and the conditions for the existence of phases. The same table also contains the order of the phase transitions between two neighboring phases. For completeness we also include in these tables the additional phases found in the planar windmill model that is discussed in Sec.~\ref{sec:xy-model}. 
\subsection{Decoupled windmill phase $(1)$}
\label{sec:deco-windm-phase}
\begin{figure}[t!]
\includegraphics[width=.8\linewidth]{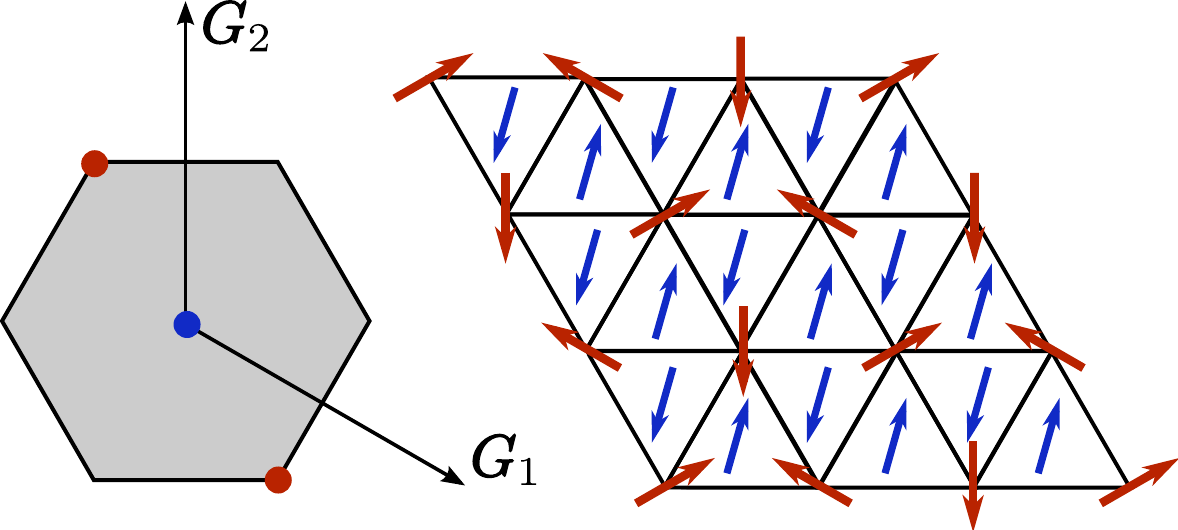}
\caption{Decoupled windmill phase ($1$). In this phase the two sublattices are decoupled. The triangular lattice spins have a 120$^\circ$ order, whereas the honeycomb spins are N\'{e}el ordered. Left: Ordering wave vectors for honeycomb (yellow) and triangular (red) lattice}
\label{phase1}
\end{figure}

In the ``decoupled windmill'' phase $(1)$ the spins on the triangular lattice are arranged in a $120^\circ$ configuration while the honeycomb spins exhibit N\'eel order. This phase exists in a large region of the phase diagram where $J_{th}\leq\sqrt{J_tJ_h}$. The spin configuration of phase $(1)$ is shown in Fig.~\ref{phase1}, and an analytical expression is given by
\begin{align}
  \label{eq:7}
{\bm S}^t ({\bm r}) &= \bmx     \cos ({{\bm Q}^t\cdot {\bm r}})\\
\sin ({{\bm Q}^t\cdot {\bm r}})\\
0 \emx \\
\label{eq:8}
{\bm S}^A({\bm r}) &= \bmx \cos ({\bm Q}^A\cdot {\bm r})\\
\sin ({{\bm Q}^A\cdot {\bm r}})\\
0 \emx \\
\label{eq:9}
{\bm S}^B({\bm r}) & = -{\bm S}^A({\bm r}) \,.
\end{align}
Here, $\bfr = n \bfa_1 + m \bfa_2$ is a Bravais lattice vector and the ordering wave vectors read in the basis of the reciprocal lattice vectors as
\begin{align}
  \label{eq:5}
  {\bm Q}^{t} &= \pm (1/3,- 1/3)\\
\label{eq:6}
{\bm Q}^{A,B} &= (0,0) \, .
\end{align}
The triangular ordering vectors are thus given by $\bfqq^t = \pm (\bfgg_1/3 - \bfgg_2/3) = \pm 2 \pi( \frac13,- \frac{1}{\sqrt{3}} )$ and are located, as shown in Fig.~\ref{phase1}, at the corners of the first Brillouin zone. Only two of those six corners are non-equivalent, \emph{i.e.}, cannot be reached by adding a reciprocal lattice vector. 
We have chosen to place the honeycomb spins in the same plane as the triangular lattice spins. This is only done for convenience, since in this configuration the spins on the triangular lattice are decoupled from the spins on the honeycomb lattice: global rotations of spins on either of the two sublattices do not cost any energy. The sign of the different wave vectors $\bfqq^t$ corresponds to different chiralities of the $120^{\circ}$-order. 

To find the order parameter manifold of this phase, we first divide the tripartite triangular lattice into ${X,Y}$ and $Z$-sites, where the different sites correspond to the three possible directions of triangular lattice spins in the $120^\circ$ configuration. Consider a plaquette of $X$-, $Y$- and $Z$-site spins and let ${\bm t}_2$ be a unit vector along the spin on the $X$-sites ${\bm S}_X$. Then define the unit vector ${\bm t}_1$ to be parallel to the component of the spin ${\bm S}_Y$ on the $Y$ sites that is orthogonal to ${\bm S}_X$. Finally, define the unit vector ${\bm t}_3={\bm t}_1\times{\bm t}_2$. The triad $({\bm t}_1,{\bm t}_2,{\bm t}_3)$ that is defined in this way is the local order parameter for the $120^\circ$ phase. The vector ${\bm t}_3$ is essential to encapsulate the chirality of the ordering, \emph{i.e.}, $XYZ$ ordering versus $XZY$. The chirality on a triangular plaquette is defined to be negative (positive), if the rotation of spins going from site $X$ to $Y$ to $Z$ is (counter)clockwise. If the plane of triangular spin order is given by the $x$-$y$-plane of real-space, as is assumed in Fig.~\ref{phase1}, a positive (negative) chirality corresponds to the unit vector $\bft_3$ pointing out-of (into) the plane. The chirality of the configuration can be changed by performing a $\pi$-rotation around an axis that lies in the plane of the triangular spins. The order parameter manifold of the triangular lattice spins is thus given by $SO(3)$. 

The order parameter of the honeycomb lattice spins can be defined by a unit vector ${\bm n}$ that points along the direction of the $A$-site spins, and global rotations around an axis that is not parallel (or anti-parallel) to $\bfn$ yield other energetically degenerate spin configurations. The order parameter manifold of the ``decoupled windmill'' phase $(1)$ is thus given by $SO(3) \times O(3)/O(2)$. 

We finally mention that at finite temperatures, it was shown in Refs.~\onlinecite{PhysRevLett.109.237205, PhysRevB.89.094417} that thermal (or quantum) fluctuations around this ground state lead to a finite temperature phase diagram with $\mathbb{Z}_6$ order and an emergent critical phase.

\subsection{Collinear antiferromagnetic phase/canted ferromagnetic phase ($2a$) and ($2b$)}
\label{sec:coll-antif-ferr}
\begin{figure}[t!]
\includegraphics[width=1.00\linewidth]{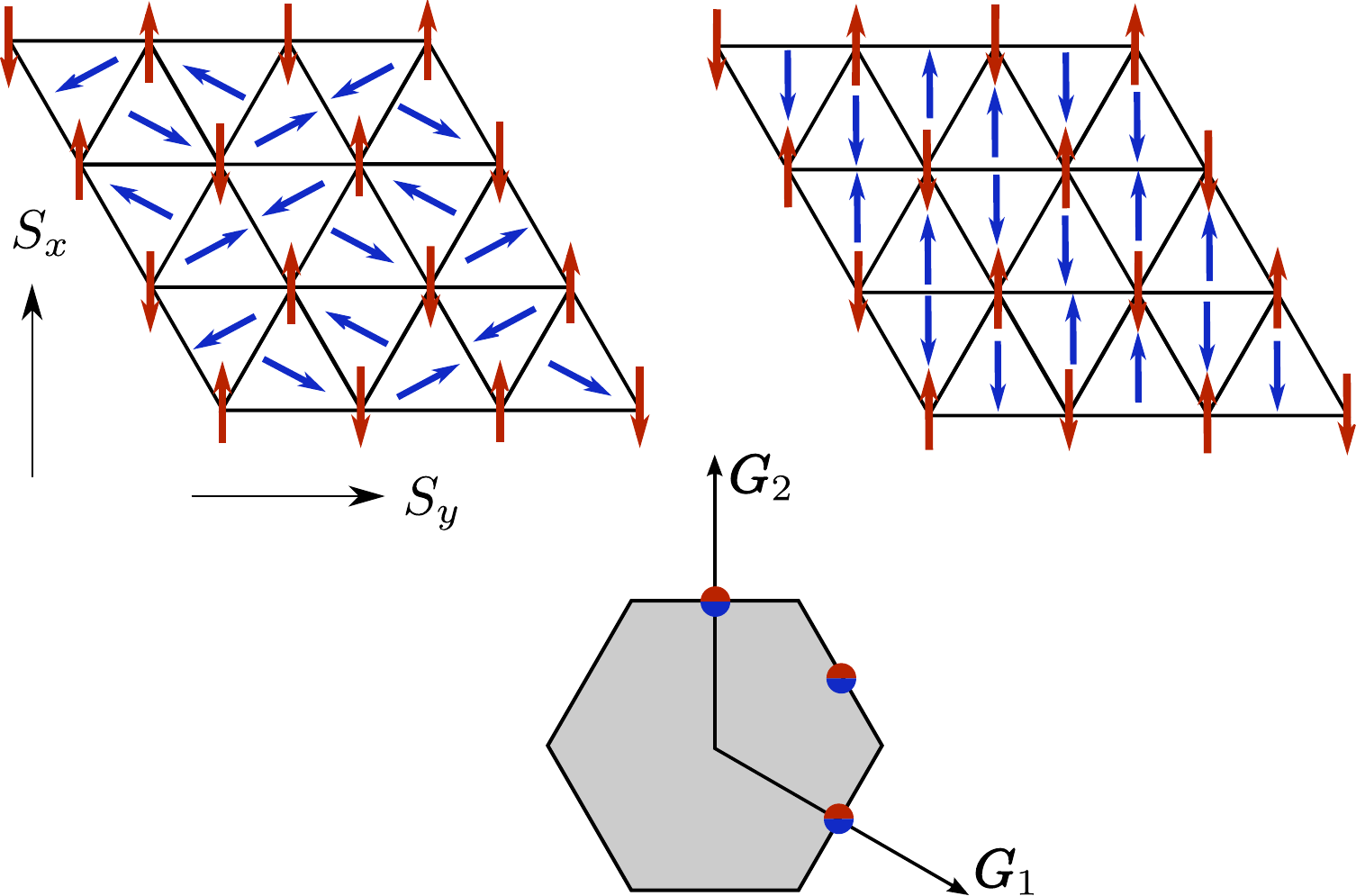}
\caption{Collinear Antiferromagnet/Canted Ferromagnet (left: $(2a)$, right: $(2b)$). Phase (2a) can exist only for $j_{th}<2j_h$, when this inequality is violated, phase (2b) minimizes the energy. In configuration (2a) the honeycomb spins can be globally rotated around the triangular spin direction without any cost in energy. The ordering wave vectors of both lattices are identical.}
\label{phase2}
\end{figure}

This is a planar phase in which the spins on the triangular sublattice are collinearly ordered, \emph{i.e.}, ferromagnetically along one lattice direction and antiferromagnetically along the others. The $A$/$B$ honeycomb sublattices are each collinear canted ferromagnets with spins on the $B$ sites that are antiparallel to two of the neighboring spins on the $A$-site. The spin configuration is shown in Fig.~\ref{phase2}. We have chosen the direction of one of the triangular lattice spins to be along the $S_x$-axis of spin space. An analytical expression of the spin configuration in this phase is given by 
\begin{eqnarray} 
{\bm S}^t ({\bm r}) &=& \left(
\begin{array}{c}
\cos ({{\bm Q}^t\cdot {\bm r}} )\\
0\\
0
\end{array}
\right)\\
{\bm S}^A({\bm r}) &=& -\left(
\begin{array}{c}
\cos (\theta +{{\bm Q}^A\cdot {\bm r}})\\
\sin \theta\\
0
\end{array}
\right)\\&=&\left(
\begin{array}{c}
-j_{th}/(2j_h)\cos({{\bm Q}^A\cdot {\bm r}})\\
-[{1-{j_{th}^2}/{(4j_h^2)}}]^{1/2}\\
0
\end{array}
\right)\\
{\bm S}^B({\bm r}) &=& -{\bm S}^A({\bm r})
\end{eqnarray}
with 
\begin{eqnarray}
\theta&=&-\cos^{-1}\left(\frac{j_{{th}}}{2j_{{h}}}\right)\\
{\bm Q}^{t}&=&{\bm Q}^{A,B} \in \{ (1/2,0),(0,1/2), (1/2, 1/2) \} \,.
\end{eqnarray}

The form of $\theta$ was obtained by taking it as a variational parameter and minimizing the energy found from the explicit evaluation of the Hamiltonian (we refer to Appendix~\ref{sec:coll-antif-ferr-1} for details).

Note that $\theta$ is undefined for $j_{{th}}>2j_{{h}}$. In this latter range the minimum is instead found at $\theta=0$. The region where this phase is hosted is denoted ($2b$) in the phase diagram. The transition between this configuration and one of finite $\theta$ is accompanied by a discontinuity of the second derivative of the energy as a function of the coupling constants, \emph{i.e.}, the phase transition between the phases $(2a)$ and $(2b)$ is of second order.  

The order parameter is given by defining the direction of one triangular lattice spin and by specifying one direction in lattice space in which the spins are collinear with the chosen spin. For the latter there are three possible choices. The honeycomb spins arrange themselves in a plane that contains the triangular spins and enclose a certain angle $\theta$ with the triangular spins. Global $O(2)$ rotations of the honeycomb spins around an axis parallel to the triangular spins yield energetically degenerate spin configurations. The order parameter manifold of the phases $(2a)$ and $(2b)$ is thus given by $O(3)/O(2)\times O(2) \times \mathbb{Z}_3$.

\subsection{Ferromagnetic phases ($3a$) and ($3b$)}
\label{sec:ferromagnet-3}
\begin{figure}[t]
\includegraphics[width=1.0\linewidth]{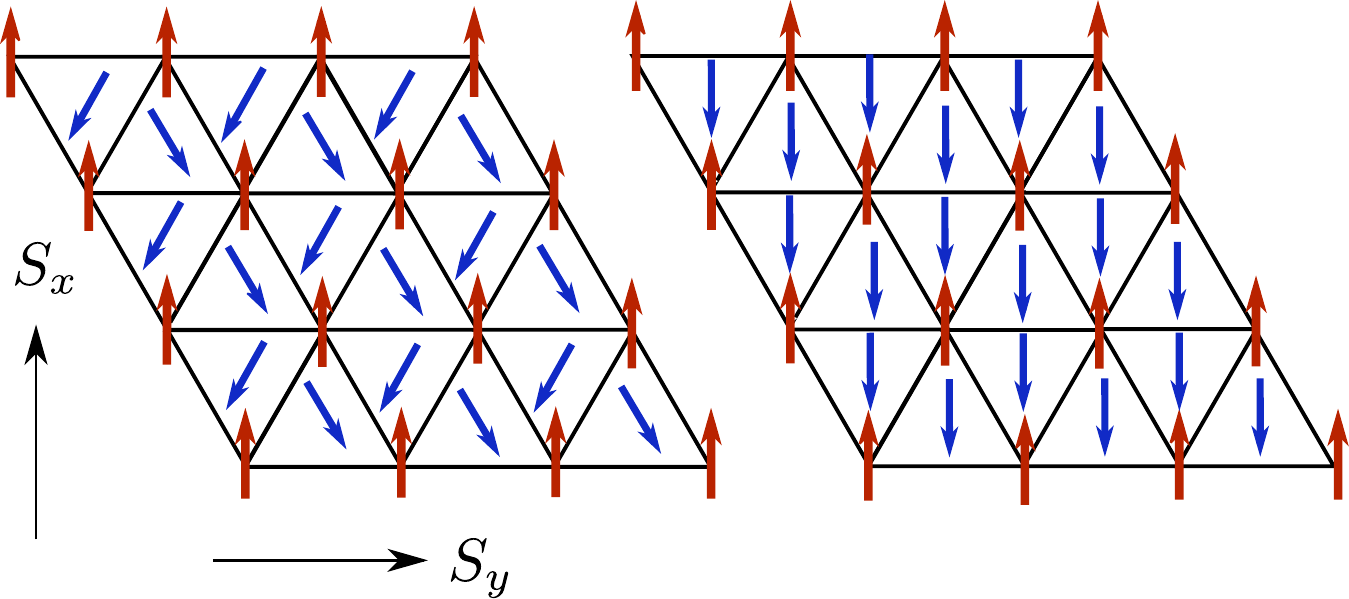}
\caption{Left: (3a), right: (3b). All three sublattices are ferromagnetically ordered. In phase (3a) the relative angles between the lattices is tunable by changing $j_{th}$. However, once $j_{th} \geq 2j_h$ the spins become locked in the configuration (3b).}
\label{phase3}
\end{figure}

In phase $(3a)$ all three sublattices are separately ferromagnetically ordered. The spins on the honeycomb $A$ and $B$ sites enclose an angle $\theta$ with the triangular spins, but point in mirror opposite directions with respect to the triangular spins. The spin configuration is depicted in Fig.~\ref{phase3}. Let the direction of ${\bm S}^t$ be the $S_x$-axis of spin space. An analytical expression of the spin configuration reads
\begin{eqnarray}
{\bm S}^t ({\bm r}) &=& \left(
\begin{array}{c}
1\\
0\\
0
\end{array}
\right)\\
{\bm S}^A({\bm r}) &=& \left(
\begin{array}{c}
\cos \theta\\
\sin \theta\\
0
\end{array}
\right)=\left(
\begin{array}{c}
-j_{th}/(2j_h)\\
\phantom{-}[{1-{j_{th}^2}/{(4j_h^2)}}]^{1/2}\\
0
\end{array}
\right)\\
{\bm S}^B({\bm r}) &=& \left(
\begin{array}{c}
\cos \theta\\
-\sin \theta\\
0
\end{array}
\right)=\left(
\begin{array}{c}
-j_{th}/(2j_h)\\
-[{1-{j_{th}^2}/{(4j_h^2)}}]^{1/2}\\
0
\end{array}
\right)
\end{eqnarray}
with 
\begin{eqnarray}
\cos(\theta)=-\frac{j_{{th}}}{2j_{{h}}}.
\end{eqnarray}
The ordering wave vectors are all identically zero, which corresponds to ferromagnetic order.
\begin{figure}[t]
\includegraphics[width=.49\linewidth]{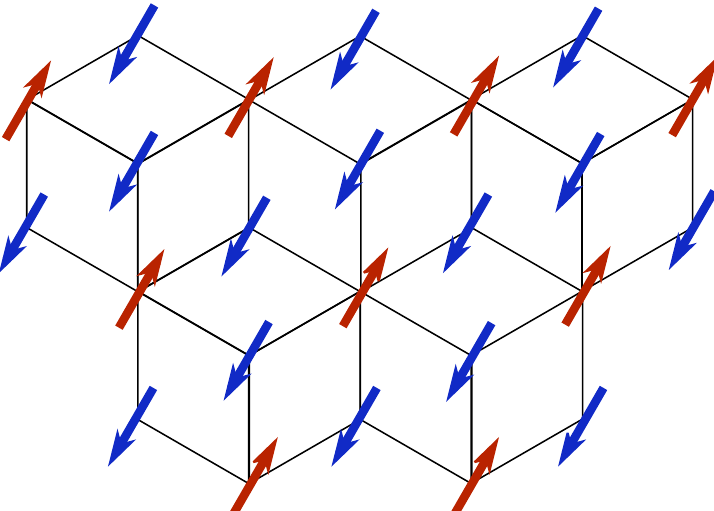}
\caption{Figure shows the \emph{dice lattice} obtained in the limit of large $J_{th} \gg J_t, J_h$. This is effectively the lattice for phase (3b). Red (blue) arrows lie on triangular (honeycomb) lattice sites. }
\label{Dice}
\end{figure}

Once again the variational parameters become undefined for $j_{{th}}>2j_{{h}}$. The minimum for this region is instead found at  $\theta=\pi$ and the corresponding phase is denoted ($3b$) in the phase diagram. This phase is the one that would be obtained in the limit of infinite $J_{th}$ and finite $J_{t},J_{h}$.  Such a limit corresponds to a lattice where there is a vanishingly small coupling within the triangular lattice and within the honeycomb lattice, but a dominant inter-sublattice coupling $J_{th}$. The resulting bipartite lattice is depicted in Fig.~\ref{Dice} and is known as the \emph {dice lattice}. It is the dual of the kagome lattice and has been studied in the literature in various contexts\cite{WangRan}. 

The order parameter manifold of this phase is defined by the direction of the triangular lattice spin. The energy of phase $(3a)$ remains invariant, however, if we rotate all honeycomb spins around an axis parallel to the triangular spin, which corresponds to an $O(2)$ symmetry. Therefore, the order parameter manifold of phase $(3a)$ is given by $O(3)/O(2) \times O(2)$, whereas that of phase $(3b)$ is equal to $O(3)/O(2)$.  

\subsection{Double cone phases ($4a$) and ($4b$)}
\label{sec:double-cone-phase}
\begin{figure}[t!]
\includegraphics[width=\linewidth]{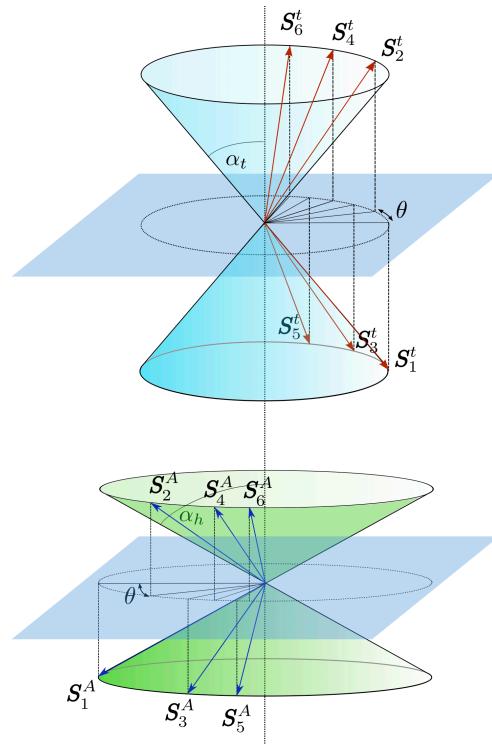}
\caption{Double Cone Phase $(4a)$. Spins on the triangular (top) and the honeycomb lattice (bottom) lie on separate double cones with different opening angles $\alpha_t$ and $\alpha_h$. In the figures we use the notation ${\bm S}_{n+1} ^{t,A,B} ={\bm S}^{t,A,B}_{1} (\bm r + n {\bm a}_1 )$ and ${\bm S}_{2n+1}^{t,A,B} ={\bm S}_1^{t,A,B} (\bm r + n {\bm a}_2 )$. }
\label{DoubleCone4a}
\end{figure}

\begin{figure}[h!]
\includegraphics[width=.8\linewidth]{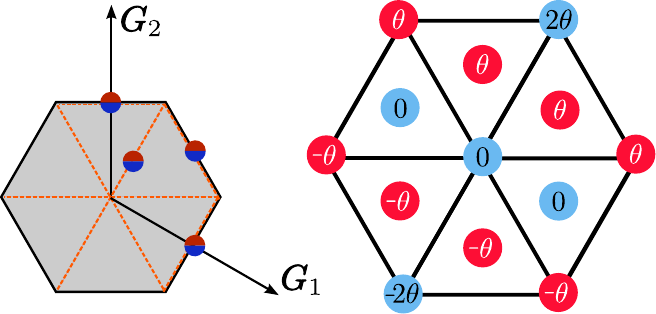}
\caption{Double Cone Phase $(4a)$. Left: $\bm Q$ vectors of all sublattices are identical. The phase contains one of the three $\bm Q$ vectors on the boundary and one from the dashed lines. Specifying $\theta$ determines the exact position of the $\bm Q$ vector on the dashed lines. Right: Scheme of spin orientations on the lattice. Let the middle reference spin be on the upper double cone of the triangular lattice. The blue (red) spins lie on upper (lower) cones. The angles describe by how much a spin's azimuthal angle differs from that of the reference spin. }
\label{Scheme4a}
\end{figure}

Spins in the phases $(4a)$ and $(4b)$ exhibit a non-coplanar arrangement. In phase $(4a)$, spins on the triangular and the honeycomb lattice lie on separate double-cones. The phase $(4a)$ is shown in Figs.~\ref{DoubleCone4a} and~\ref{Scheme4a}. The double cone structure is depicted in Fig.~\ref{DoubleCone4a} and the arrangement of the spins on the lattice is clearly illustrated in Fig.~\ref{Scheme4a}. As one moves horizontally in lattice space along $\bfa_1$ ($\bfa_2$) from one site to the next, the spins advance by an azimuth angle of $\theta$ $(2 \theta)$. In other words, the spins in the figure labelled by the same angles share a common plane in spin space. Additionally, they alternately lie on upper and lower cones, which is represented by blue and red in Fig.~\ref{Scheme4a}. We are showing only one of the six symmetry-related ground states. The others are obtained by rotating the lattice through an angle of $(2\pi/6) n$ with $n=1,\dots, 5$. 

Non-coplanar ground states, similar to the ones we have found, have also been discovered in Heisenberg models on triangular, square, pyrochlore and octahedral lattices~\cite{PhysRevLett.105.047203,PhysRevLett.113.087204,PhysRevB.88.024407,LapaHenley-arXiv-2012}. The interest in such phases stems from the fact that non-coplanar spin orderings are expected to give rise to an anomalous Hall effect\cite{Taguchi}, due to the non-vanishing spin chirality. Moreover, considering quantum fluctuations it is expected that non-coplanar classical ground states give rise to chiral spin liquid phases~\cite{sachdev_qpt_book,PhysRevB.88.024407,PhysRevLett.108.207204}.

To describe this configuration we choose the opening axis of the cones to be the $S_z$-axis. Then the spins are described by
\begin{eqnarray}
  \label{eq:12}
  {\bm S}^t ({\bm r}) &=& \bmx \sin (\alpha_t) \cos ({\bm Q}_1\cdot {\bm r})\\
\sin (\alpha_t) \sin ({\bm Q}_1\cdot {\bm r})\\
\cos (\alpha_t) \cos ({\bm Q}_2\cdot {\bm r}) \emx
\\
\label{eq:13}
{\bm S}^A({\bm r}) &=& \bmx -\sin (\alpha_{{h}}) \cos ({\bm Q}_1\cdot {\bm r}+\theta)\\
-\sin (\alpha_{{h}}) \sin ({\bm Q}_1\cdot {\bm r}+\theta)\\
-\cos (\alpha_{{h}}) \cos ({\bm Q}_2\cdot {\bm r})
\emx \\
{\bm S}^B({\bm r}) &=&{\bm S}^A({\bm r}+{\bm a_1})  \label{eq:14}
\end{eqnarray}
with ordering wave vectors
\begin{eqnarray} 
{\bm Q}_1 &=& \pm\frac{1}{2\pi}(\theta,2\theta)\\
{\bm Q}_2&=&(1/2,0) \label{symmetry4a}
\end{eqnarray}
The ambiguity in sign corresponds to two possible orderings that are related by a $\pi$-rotation in lattice space. Four other orderings are possible, that correspond to the remaining rotated states
\begin{eqnarray} 
{\bm Q}_1 &=& \pm\frac{1}{2\pi}(\theta,-\theta)\\
{\bm Q}_2&=&(1/2,1/2)
\end{eqnarray}
and 
\begin{eqnarray} 
{\bm Q}_1 &=& \pm\frac{1}{2\pi}(2\theta,\theta)\\
{\bm Q}_2&=&(0,1/2).
\end{eqnarray}

The angles $\alpha_t$ and $\alpha_{{h}}$ are the half-opening angles of the respective cones. The angle $\theta$ is the difference in the azimuthal angle, as shown in Fig.~\ref{DoubleCone4a}. These three angles $\{\alpha_t, \alpha_h,\theta\}$ are functions of the coupling constants. By solving the minimization problem (see Appendix~\ref{sec:double-cone-conf}) for the set of variables $\{\alpha_t, \alpha_h,\theta\}$, one finds the solution
\begin{eqnarray}
\sin \alpha_t &=& \sqrt{\frac{1-\sigma^2}{\rho^2-\sigma^2}}\label{alpha_t}\\
\sin \alpha_h &=& \sqrt{\frac{1-\sigma^2}{\rho^2-\sigma^2}}\rho \label{alpha_h}\\
\cos \theta &=& j_{{th}} \frac{j_{{th}}-\sqrt{j_{{th}}^2-4j_{{h}}}}{4j_{{h}}}\label{theta}=\frac{j_{{th}}}{2}\rho
\end{eqnarray}
with
\begin{eqnarray}
\rho \equiv \frac{\sin \alpha_{{h}}}{\sin \alpha_t}&=&\frac{j_{{th}}-\sqrt{j_{{th}}^2-4j_{{h}}}}{2j_{{h}}} \label{rho}\\
\sigma \equiv \frac{\cos \alpha_{{h}}}{\cos \alpha_t}&=&\frac{j_{{th}}\left(j_{{th}}-\sqrt{j_{{th}}^2-4j_{{h}}}\right)^2}{8j_{{h}}^2} \label{sigma}.
\end{eqnarray}
Generally both angles $\alpha_t, \alpha_h$ are nonzero. 

\begin{figure}
\includegraphics[width=.9\linewidth]{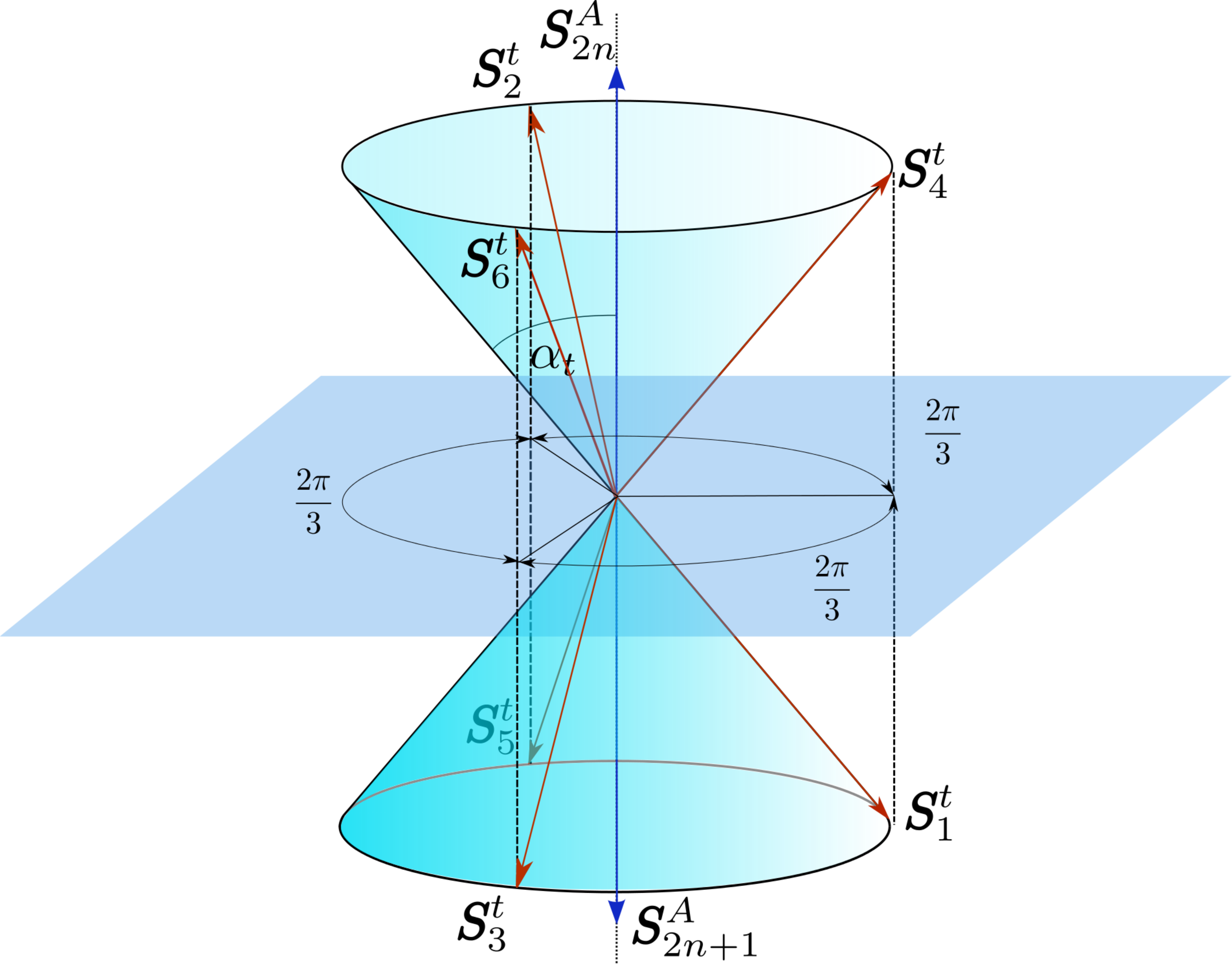}
\caption{Double Cone Phase $(4b)$. The spins on the honeycomb lattice align with the cone axis, while the spins on the triangular lattice lie on a double cone with half-opening angle $\cos{\alpha_t}=2j_{th}$ and advance by a fixed angle of $\theta = \frac{2 \pi k}{3}$ when translated by $\bfa_k$. }
\label{DoubleCone4b}
\end{figure}

Another solution to the minimization problem exists, for which $\rho=0$. The corresponding phase is denoted by $(4b)$ and is shown in Fig.~\ref{DoubleCone4b}. In this configuration the honeycomb lattice spins are aligned with the cone axis, while the triangular lattice spins lie again on a cone with variable half-opening angle
\begin{eqnarray}
\cos\alpha_t=2j_{th} \,.
\end{eqnarray}
Consecutive triangular lattice spins rotate by an angle of ${\theta=2\pi/3}$ around the cone axis. 

The order of both phases may be described by first specifying the cone axis  and one of the triangular lattice spins. Once this choice is made, there is a further degeneracy due to the freedom in rotating the lattice, as explained above. Thus there are six energetically degenerate configurations. These six possibilities may be further classified into two chiralities: three cover the configurations that describe a right handed screw around the cone axis when advancing in the $\bm a_1$ direction and three describe a left handed screw. In specifying the cone axis, we may encode the chirality of the spins via the direction of the cone axis vector. The chirality can thus be reversed by a $\pi$-rotation around an axis that lies in the plane orthogonal to the cone axis. The order in this phase is therefore described by an element of $SO(3)\times {\mathbb Z}_3$.

\subsection{Incommensurate spiral phase $(5)$}
\label{sec:incomm-spir-5}
\begin{figure}[t]
\includegraphics[width=\linewidth]{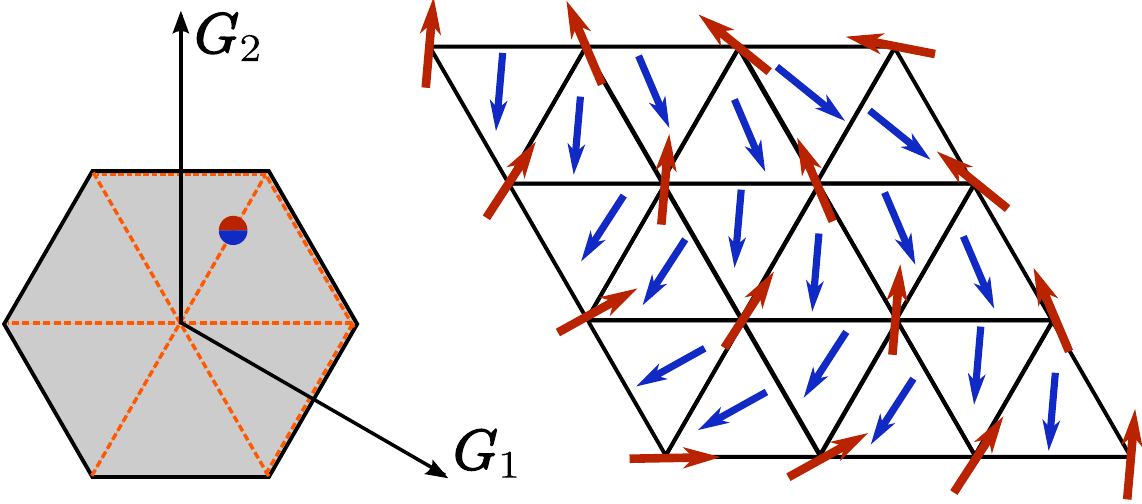}
\caption{Phase $(5)$. Left: $\bm Q$ vector is identical for all sublattices. It lies on one of the dashed lines and is determined by $\theta$.  Right: The spins exhibit incommensurate spiral order. The spiral can be either left-handed or right-handed.}
\label{phase5}
\end{figure}

In phase $(5)$ the spins are arranged in a coplanar incommensurate spiral as shown in Fig.~\ref{phase5}. The spin configuration takes the form
\begin{eqnarray}
{\bm S}^t ({\bm r}) &=& \left(
\begin{array}{c}
\cos ({\bm Q}\cdot {\bm r}) \\
\sin ({\bm Q}\cdot {\bm r})\\
0
\end{array}
\right)\\
{\bm S}^A({\bm r}) &=& -\left(
\begin{array}{c}
\cos\left( {\bm Q}\cdot{\bm r}+\theta \right)\\
\sin \left( {\bm Q}\cdot{\bm r}+\theta \right)\\
0
\end{array}
\right)\\
{\bm S}^B({\bm r}) &=& -\left(
\begin{array}{c}
\cos\left( {\bm Q}\cdot{\bm r}+2\theta \right)\\
\sin \left( {\bm Q}\cdot{\bm r}+2\theta \right)\\
0
\end{array}
\right) \,,
\end{eqnarray}
where the ordering wave vector is given by 
\begin{eqnarray} 
{\bm Q} &=&\pm\frac{1}{2\pi} (\theta,2\theta).
\end{eqnarray}
with the angle $\theta$ defined via
\begin{align}
  \label{eq:10}
  \cos \theta = -\frac{1}{2} \left(1+j_{ h}-2 j_{ th}\right)\,.
\end{align}
As for the double cone phase, the sign ambiguity of $\bfqq$ describes the chirality. Four further symmetry related configurations exist and are described by the $\bm Q$ vectors
\begin{eqnarray}
{\bm Q} &=&\pm\frac{1}{2\pi} (2\theta,\theta)
\end{eqnarray}
and
\begin{eqnarray}
{\bm Q} &=&\pm\frac{1}{2\pi} (\theta,-\theta).
\end{eqnarray}
The right-hand side of Eq.~\eqref{eq:10} must satisfy
\begin{eqnarray}
\left| \left(1+j_{ h}-2 j_{ th}\right)\right| \leq 2 \,.
\end{eqnarray}
For parameters that violate this inequality the minimum is instead found at either $\theta=0$ or $\theta=\pi$. The former case is identical to phase $(3b)$ and the latter one to phase $(2b)$.

In the spiral phase the triangular lattice spins have the same order parameter manifold as the $120^\circ$ ordered spins of phase $(1)$. This is clear, since the $120^\circ$ order is just a special case of this incommensurate spiral phase. An incommensurate spiral phase on the windmill lattice, however, can appear in three distinct types, as described above, three of each chirality. The order parameter manifold of the incommensurate spiral phase is thus equal to $SO(3)\times{\mathbb Z}_3$.

\section{Planar (XY) windmill model}
\label{sec:xy-model}
\begin{figure}
\includegraphics[width=\linewidth]{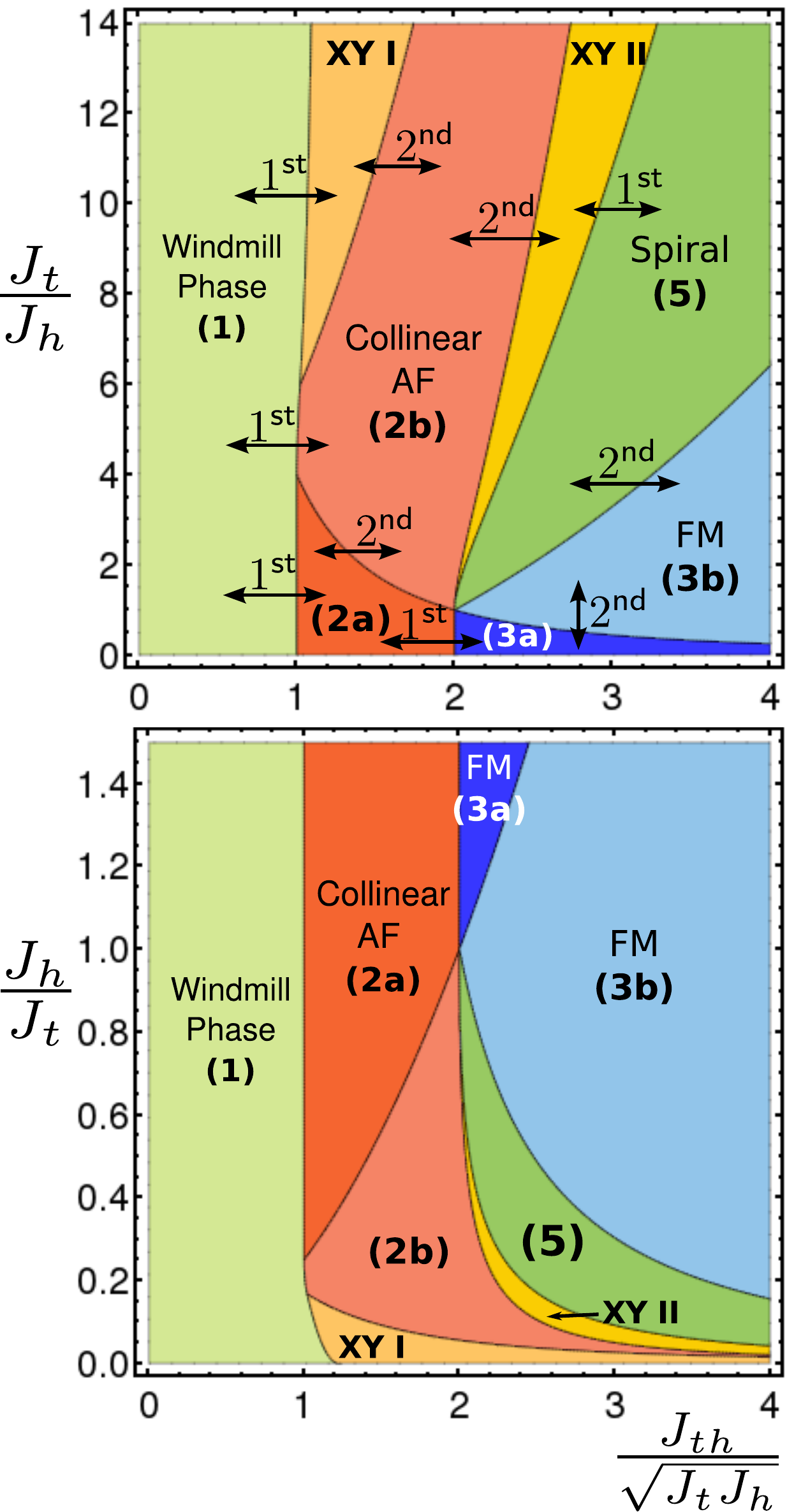}
\caption{Ground state phase diagram of the XY windmill model. The top figure has as the vertical axis $J_{t}/J_h$, while in the lower panel it is $J_h/J_t$ . The $6$ planar phases of the Heisenberg windmill model reappear, with two additional phases labelled ($\text{XY I}$) and ($\text{XY II}$)} 
\label{phasediag}
\end{figure}

\begin{table}[b!]
\centering
\begin{tabular}{c|c}
\hline\hline
{Phase} & Order Parameter Manifold (planar model)\\
\hline
1 & $O(2)\times SO(2)$ \\
2a, 2b & $SO(2) \times \mathbb{Z}_6$\\
3a & $SO(2)\times {\mathbb Z}_2$ \\ 
3b &  $SO(2)$ \\ 
5 & $ O(2)\times {\mathbb Z}_3$   \\
XY I   & $O(2)\times {\mathbb Z}_3$ \\
XY II & $O(2)\times {\mathbb Z}_3$\\
\hline\hline
\end{tabular}
\caption{Order parameter manifolds in the planar windmill model.}
\label{tablesymm}
\end{table}

We now discuss the XY windmill model, where planar (XY) spins $\bfss^\alpha_i = (S^\alpha_{i,x}, S^\alpha_{i,y})$, $\alpha \in \{t, A, B\}$, are placed on the vertices of the windmill lattice (see Fig.~\ref{model}). This case is particularly interesting given the recent advances in the field of ultracold atoms, where XY spins with nearest-neighbor antiferromagnetic interactions were successfully simulated on a triangular lattice~\cite{Struck19082011,StruckSengstockMathey-NatPhys-2013}. The lattice is created by means of standing wave laser fields. The atoms are trapped in and tunnel between the local minima of this optical lattice. At low temperatures, the system becomes superfluid and the atoms have well-defined local condensate phases $\phi_i$ at every site $i$ of the lattice. This U(1) degree of freedom plays the role of the XY spins $\bfss_i = (\cos \phi_i, \sin \phi_i)$. The nearest-neighbor spin couplings are determined by the tunneling amplitudes of atoms that move between the different laser field minima. For normal quantum mechanical tunneling between the sites, the corresponding spin interaction is always ferromagnetic, favoring a locking of the condensate phases to the same value. In contrast, it was recently demonstrated that the tunneling element acquires a non-zero Peierls phase by periodically shaking the optical lattice and an antiferromagnetic coupling of local XY phases on a triangular lattice was experimentally realized~\cite{Struck19082011,StruckSengstockMathey-NatPhys-2013}. As honeycomb optical lattice geometries have also been successfully implemented in various groups~\cite{SoltanPanahiSengstock-NaturePhys-2011,EsslingerDirac-Nature2012,arxiv:1406.7874}, it is entirely feasible to realize the XY windmill model using cold bosonic atoms in optical lattices.

We have determined the full phase diagram of the XY windmill model. It is shown in Fig.~\ref{phasediag}. We note that a coplanar state that minimizes the energy for Heisenberg spins has to minimize the energy for XY spins, as well. Thus, compared to the results for the Heisenberg windmill model the only major modification in the phase diagram, apart from a quantitative shift of the phase boundaries, takes place in the regions where the Heisenberg model exhibits non-coplanar phases. These regions are now largely occupied by two new planar phases, (XY I) and (XY II), which we describe below. 

Of the two highly degenerate points in the phase diagram of the Heisenberg model, only one remains in the planar model: at the point $(j_h, j_{th}/\sqrt{j_h}) = (1, 2)$ six phases meet.  Like in the case of the Heisenberg model, this high degeneracy is explained by the fact that all six phases deform into either phase $(2b)$ or phase $(3b)$ in the limit $(j_h, j_{th}/\sqrt{j_h}) \rightarrow (1, 2)$.  Thus there are only two distinct phases present at this point. In particular, phase (XY II) transforms into phase $(2b)$, as we explain below.

\subsection{Incommensurate alternating spiral phase (XY I)}
\label{sec:xy-i}
\begin{figure}
\includegraphics[width=\linewidth]{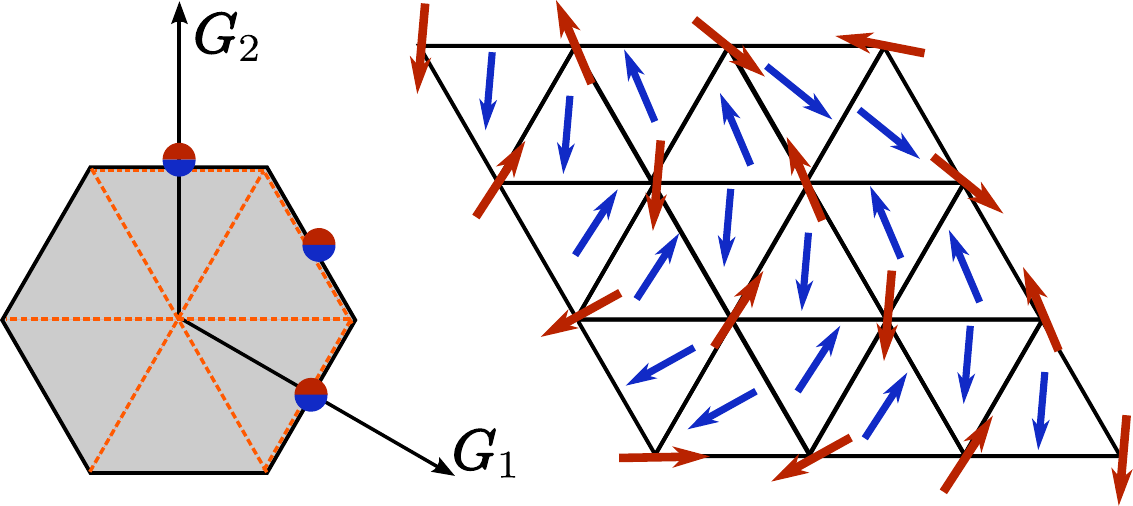}
\caption{Phase (XY I). Left: $\bm Q$ vectors are identical for all sublattices. By specifying $\theta$ the wave vectors come to lie either on the boundary or on the dashed lines. Right: The spins arrange themselves in an incommensurate, alternating spiral.}
\label{phasexy1}
\end{figure}

In place of the non-coplanar phase $(4b)$ we now find a new planar phase. We denote this phase by (XY I). It is an incommensurate spiral phase with the directions of spins alternating from one site to the next. It is shown in Fig.~\ref{phasexy1}. It may be described as a twisted N\'{e}el-ordered configuration. The spin configuration is given by
\begin{eqnarray}
{\bm S}^t ({\bm r}) &=& \left(
\begin{array}{c}
\cos ({\bm Q}\cdot {\bm r}) \\
\sin ({\bm Q}\cdot {\bm r})
\end{array}
\right)\\
{\bm S}^A({\bm r}) &=& \left(
\begin{array}{c}
\cos\left( {\bm Q}\cdot{\bm r}+\theta \right)\\
\sin \left( {\bm Q}\cdot{\bm r}+\theta \right)
\end{array}
\right)\\
{\bm S}^B({\bm r}) &=& \left(
\begin{array}{c}
\cos\left( {\bm Q}\cdot{\bm r}+2\theta \right)\\
\sin \left( {\bm Q}\cdot{\bm r}+2\theta \right)
\end{array}
\right)
\end{eqnarray}
with ordering wave vectors
\begin{eqnarray} 
{\bm Q} &=& \pm\frac{1}{2\pi} (\theta+\pi,2\theta)\,,
\end{eqnarray}
and an angle $\theta$ that is defined by 
\begin{align}
  \label{eq:11}
  \cos \theta = \frac{1}{2} \left(1+j_{ h}+2 j_{ th}\right)\,.
\end{align}
The chiralities are encoded in the sign of $\bm Q$. Four further symmetry related phases are possible with $\bm Q$ vectors
\begin{eqnarray} 
{\bm Q} &=& \pm\frac{1}{2\pi} (2\theta,\theta+\pi)
\end{eqnarray}
and
\begin{eqnarray} 
{\bm Q} &=& \pm\frac{1}{2\pi} (\theta+\pi,\pi-\theta).
\end{eqnarray}
Imposing the condition that the angle $\theta$ be real, yields
\begin{eqnarray} 
j_h+2 j_{th}\leq 1.
\end{eqnarray}
This phase is never an energy minimum for the Heisenberg model, since the non-coplanar phase (4b) includes the whole range of existence of phase (XY I) and has a smaller energy. 

Like in the case of the incommensurate spiral phase $(5)$, the order parameter of phase (XY I) is determined by giving the direction of two triangular lattice spins and by specifying one of the three $\bm Q$ vectors. Hence, its manifold is $O(2)\times {\mathbb Z}_3$.

\subsection{Canted ferromagnetic phase (XY II)}
\label{sec:xy-ii}
\begin{figure}
\includegraphics[width=.6\linewidth]{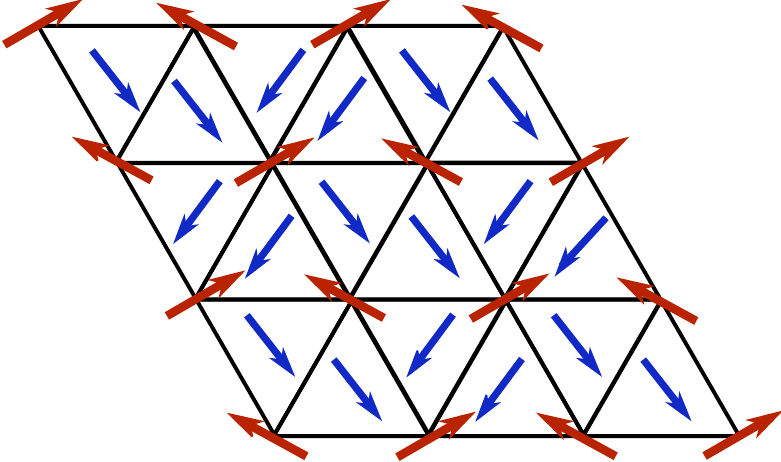}
\hspace{.02\linewidth}
\includegraphics[width=.3\linewidth]{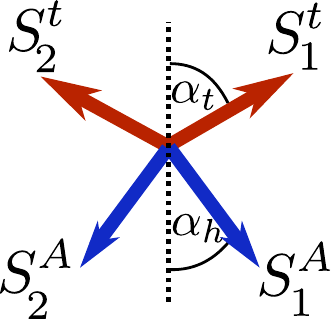}
\caption{Left: Phase (XY II). The spins on all three sublattices alternate between two positions. These two positions may be viewed as reflections along a mirror line that is common to all three sublattices. Right: Shown are spins from the triangular lattice (red) and ($A$-site) honeycomb lattice (blue) at neighboring lattice sites along $\bfa_1$. The horizontal line is a mirror line, analogous to the mirror plane of the double cone phase. The spins are aligned ferromagnetically along $\bfa_2$.}
\label{xy2}
\end{figure}

The phase (XY II) is found in the former region of the non-coplanar phase $(4a)$ in the phase diagram. It is depicted on the left of Fig.~\ref{xy2}. All spins lie symmetrically about a line, which we choose in the following to be the $x$-axis. 

The spin configuration is related to the double-cone state, since it can be obtained by forcing the double-cone azimuth angle $\theta$ to be zero (see Eqs.~\eqref{eq:12}-\eqref{eq:14}). 
The spin configuration is given by
\begin{eqnarray}
{\bm S}^t ({\bm r}) &=& \left(
\begin{array}{c}
\cos ({\alpha_t}) \\
\sin ({\alpha_t+\bm Q}\cdot {\bm r})
\end{array}
\right)\\
{\bm S}^A({\bm r}) &=& \left(
\begin{array}{c}
-\cos (\alpha_h) \\
-\sin \left(\alpha_h+ {\bm Q}\cdot{\bm r}\right)
\end{array}
\right)\\
{\bm S}^B({\bm r}) &=& \left(
\begin{array}{c}
-\cos (\alpha_h) \\
\sin \left(\alpha_h+ {\bm Q}\cdot{\bm r}\right)
\end{array}
\right)
\end{eqnarray}
with 
\begin{eqnarray} 
{\bm Q} &\in& \left\{(1/2,0),(0,1/2),(1/2,1/2)\right\}.
\end{eqnarray}
The angles $\alpha_h$ and $\alpha_t$ are given by
\begin{eqnarray}
\sin \alpha_t &=& \sqrt{\frac{1-\sigma^2}{\rho^2-\sigma^2}}\label{alpha_t}\label{xy2alpha}\\
\sin \alpha_{{h}} &=& \sqrt{\frac{1-\sigma^2}{\rho^2-\sigma^2}}\rho \label{alpha_h}
\end{eqnarray}
with
\begin{align}
    \label{eq:15}
    \sigma &\equiv \frac{\cos \alpha_{{h}}}{\cos \alpha_t} \\
&=\frac{2}{3j_{th}} + \frac{j_{th}}{3j_h} - \sqrt{\left(\frac{2}{3j_{th}} + \frac{j_{th}}{3j_h}\right)^2-\frac{1}{j_h}} \nonumber \\
\rho &\equiv \frac{\sin \alpha_{{h}}}{\sin \alpha_t}=3\sigma-\frac{4}{j_{th}}\,. \label{xy2rho}
  \end{align}
As we approach the high degeneracy point $(j_h, j_{th}/\sqrt{j_h}) = (1, 2)$, it is straightforward to show from (\ref{xy2alpha})-(\ref{xy2rho}) that  $\alpha_h,\alpha_t \rightarrow \pi/2$ as the degeneracy point is approached and (XY II) becomes identical to phase $(2b)$. 

Concerning the lattice symmetry of this phase, we note that from the configuration shown in Fig. \ref{xy2} two further energetically degenerate phases are obtained by rotation of the lattice by (i) $120^\circ$ and (ii) $240^\circ$, respectively. Hence, the order parameter manifold in this phase is equal to $O(2)\times {\mathbb Z}_3$.

\section{Conclusions}
\label{sec:conclusions}
We have determined the complete ground state phase diagram of both the classical Heisenberg and the planar (XY) spin models on the windmill lattice. Like the well-known $J_1$-$J_2$-model on the square lattice, the windmill model couples a lattice to its dual lattice. In the windmill model this is a triangular lattice coupled to its dual honeycomb lattice. Competing antiferromagnetic interactions between the spins lead to a rich ground state phenomenology with collinear, coplanar, incommensurate spiral and non-coplanar phases. We discussed different routes to an experimental realization of these spin models. 
Based on our results and recent finite temperature studies of the Heisenberg windmill model in the regime of weak sublattice coupling, the region of phase ($1$), we expect interesting physics to emerge in the presence of quantum and thermal fluctuations. 

\acknowledgements
We acknowledge useful discussions with P. Chandra, P. Coleman, R. Flint, D. Mendler, K. A. Ross, C. Seiler, and J. Schmalian. The Young Investigator Group of P.P.O. received financial support from the ``Concept for the Future'' of the KIT within the framework of the German Excellence Initiative.

\appendix
\section{Energies}
\label{sec:energies}
For all the phases that we have found, the spin configurations are such that the interaction energy of a spin with its neighbors is translationally invariant, \emph{i.e.}, the nearest-neighbor sums
\begin{eqnarray}
\sum_j {\bm S}^t_i \cdot {\bm S}^t_j, \sum_j {\bm S}^A_i \cdot {\bm S}^B_j, \sum_j {\bm S}^{t}_i \cdot {\bm S}^{A/B}_j
\end{eqnarray}
are all independent of $j$. As a consequence, the computation of the total energy is rather straightforward. One merely needs to consider the energy of a spin and its neighbors and multiply the result by $N/2$ (for spins on the triangular lattice) or $N$ (for spins on the honeycomb lattice). In some of the phases, however, there are configuration parameters that change continuously with the coupling constants. We introduce these as variational parameters that have to be chosen appropriately in order to minimize the total energy.

\subsection{Collinear antiferromagnetic phase/canted ferromagnetic phase $(2a)$  and $(2b)$}
\label{sec:coll-antif-ferr-1}
To calculate the energy of these phases, consider the left configuration in Fig.~\ref{phase2}. Out of the neighboring $6$ honeycomb spins of a triangular spin there are always two spins that have the same orientation (positive scalar product) as the triangular spin. One is an A-site spin, the other a B-site spin. All remaining honeycomb spins are either equal or opposite to one of the two. We introduce as the variational parameters $\theta_1$ and $\theta_2$, the angles that the A- and B-site honeycomb spins make with the triangular lattice spin. For the energy we find
\begin{eqnarray}
E_{2}&=&E_{t}+E_{h}+E_{th}\\
&=&-NJ_{t}-2NJ_{h}+NJ_{h}\cos(\theta_{1}+\theta_{2})\\
& & -NJ_{th}\left(\cos\theta_{1}+\cos\theta_{2}\right)
\end{eqnarray}
which is minimized by 
\begin{align}
  \label{eq:17}
\theta\equiv\theta_1 =\theta_2= \cos^{-1} \Bigl( \frac{J_{th}}{2J_h} \Bigr)
\end{align}
with energy
\begin{eqnarray}
E_{2a}=-NJ_{t}-3NJ_{h}-N\frac{J_{th}^{2}}{2J_{h}}.
\end{eqnarray}
In order for Eq.~\eqref{eq:17} to be meaningful, it is required that $J_{th}\leq 2J_h$. In the opposite regime $J_{th}\geq2J_h$, we instead have the minimum at $\theta_1= \theta_2=0$. The energy for this locked phase is
\begin{eqnarray}
E_{2b}=-NJ_{t}-NJ_{h}-2N J_{th}. 
\end{eqnarray}

\subsection{Ferromagnetic phases $(3a)$ and $(3b)$}
\label{sec:ferr-phase-3}
To compute the energy of this phase, we note from Fig.~\ref{phase3} that the A and B sublattices are each ferromagnetically ordered. Let $\theta_1$ and $\theta_2$ denote the angles between A and triangular sublattice magnetizations and between B and the triangular lattice magnetization, respectively. The total energy in this phase is
\begin{align}
E&= 3NJ_{t}+3NJ_{h}\cos(\theta_{1}+\theta_{2})  \nonumber \\
&\quad +3NJ_{th}\left(\cos\theta_{1}+\cos\theta_{2}\right) \,. 
\label{en3} 
\end{align}
Minimization yields the angles
\begin{eqnarray}
\cos \theta\equiv \cos \theta_{1} & = &\cos \theta_2=-\frac{j_{th}}{2j_h}
\end{eqnarray}
and energy 
\begin{eqnarray}
E_{3a} & = & 3NJ_{t}-3NJ_{h}- N\frac{3 J_{th}^{2}}{2J_{h}}.
\end{eqnarray}
The coupling constants have to satisfy 
\begin{eqnarray}
\frac{J_{th}}{2J_h}\leq 1
\end{eqnarray}
in order for $\theta$ to be real.

When this condition is violated, the minimum of the energy function is instead found at 
\begin{eqnarray}
\theta=\theta_1=\theta_2=\pi
\end{eqnarray}
with
\begin{eqnarray}
E_{3b} = 3NJ_{t}+3NJ_{h}-6NJ_{th}.
\end{eqnarray}
This is phase $(3b)$ in the phase diagram.

\subsection{Double Cone Configuration $(4a)$ and $(4b)$}
\label{sec:double-cone-conf}
We parametrize the energy of this phase in terms of the cone opening angles $\alpha_{h}$ and $\alpha_{t}$ as well as the advance angle $\theta$. Let the $z$ axis be parallel to the cone axis. Then we can write the total energy as 
\begin{align}
  \label{eq:19}
E_4 & =   -NJ_t-NJ_{h}+ 2NJ_t\sin^{2}\alpha_{t}\bigl(\cos^{2}\theta+\cos\theta\bigr) \nonumber \\
&  +2NJ_{h}\sin^{2}\alpha_{h}\bigl( \cos\theta+1\bigr) -2NJ_{th}\Bigl(\cos\bigl(\alpha_{h}-\alpha_{t}\bigr) \nonumber \\ 
& +2\sin\alpha_{t}\sin\alpha_{h}\cos\theta \Bigr)\,.
  \end{align}
Minimizing with respect to $\{\alpha_t, \alpha_h, \theta\}$ yields the three equations
\begin{align}
  \label{eq:20}
  2(\cos^2 \theta +\cos \theta)+j_{th} (\sigma-\rho-2\rho \cos \theta) &= 0 \\
\label{eq:21}
2j_h (\cos \theta+1)\sigma \rho +j_{th} (\rho-\sigma-2\sigma \cos \theta)&=0 \\
\label{eq:22}
2\cos \theta +1+j_h \rho^2- 2j_{th} \rho &= 0 \,.
\end{align}
We have introduced the definitions
\begin{align}
  \label{eq:23}
\rho &= \frac{\sin\alpha_h}{\sin\alpha_t}\\
  \label{eq:24}
\sigma&= \frac{\cos\alpha_h}{\cos \alpha_t} \,.
\end{align}
In order to find the solutions of this set of equations, we solve Eq.~\eqref{eq:22} for $\cos \theta$ and substitute this solution into the other two equations~\eqref{eq:20} and~\eqref{eq:21}
\begin{align}
  \label{eq:25}
-1 - 2 j_h j_{th} \rho^3 + j_h^2 \rho^4 + 2 j_{th} \sigma &= 0 \\
\label{eq:26}
\rho \left(j_{th} - 2 j{th}^2 \sigma + 3 j_h j_{th} \rho \sigma +j_h (1 - j_h \rho^2) \sigma\right) &= 0  
\end{align}
One obvious solution comes from the second equation with $\rho=0$. We will treat this case later. We assume $\rho\neq0$, then the two resulting equations contain $\sigma$ linearly and eliminating it results in a $6$th order equation for $\rho$ that happens to be solvable and has the roots
\begin{eqnarray}
\rho_1&=&\frac{j_{th}-\sqrt{j_{th}^2-4j_h}}{2j_h}\\
\rho_2&=&\frac{j_{th}+\sqrt{j_{th}^2-4j_h}}{2j_h}\\
\rho_3&=&\frac{j_{th}-\sqrt{j_h+j_{th}^2}}{j_h}\\
\rho_4&=&\frac{j_{th}+\sqrt{j_h+j_{th}^2}}{j_h} \,.
\end{eqnarray}
The last two solutions $\rho_3$ and $\rho_4$ are both doubly degenerate. From these solutions the resulting values for $\sigma$ and $\theta$ are found to be
\begin{eqnarray}
\sigma_1&=&\frac{j_{th} (j_{th}-\sqrt{j_{th}^2-4j_h})^2}{8j_h^2}\\
\sigma_2&=&\frac{j_{th} (j_{th}-\sqrt{j_{th}^2-4j_h})^2}{8j_h^2}\\
\sigma_3&=&\frac{-j_{th}+\sqrt{j_h+j_{th}^2}}{j_h}\\
\sigma_4&=&-\frac{j_{th}+\sqrt{j_h+j_{th}^2}}{j_h}
\end{eqnarray}
and 
\begin{eqnarray}
\cos \theta_1&=&j_{th} \frac{j_{th}-\sqrt{j_{th}^2-4j_h}}{4j_h}\\
\cos \theta_2&=&j_{th} \frac{j_{th}+\sqrt{j_{th}^2-4j_h}}{4j_h}\\
\cos \theta_3&=&-1\\
\cos \theta_4&=&-1.
\end{eqnarray}
Note that $\cos \theta_2\geq 1$, thus the solution $\theta_2$ is not admissible. The solution $\theta_3$ may also be discarded, because it has a negative value of $\rho$, whereas $\rho \geq 0$ from the definition and $0 \leq \alpha_h, \alpha_t \leq \pi$. 

The solution $\theta_4$  has $\rho_4=-\sigma_4$. Inserting into this relation the definitions in terms of $\alpha_t$ and $\alpha_h$ we have
\begin{eqnarray}
\sin \alpha_h = \rho_4 \sin \alpha_t \\
\cos \alpha_h = -\rho_4 \cos \alpha_t 
\end{eqnarray}
and by taking squares and adding, $\rho_4=1$ is deduced and as a consequence $\alpha_t=\pi-\alpha_h$. Such a phase, however, has energy $E=-NJ_t-NJ_h+2NJ_{th}$. This energy is always larger than that of phase $(1)$. We can therefore discard this solution as well.

To summarize, the only solution that needs to be considered is $(\rho_1,\sigma_1,\theta_1)$ and we therefore drop the subscript in the following. We can express the angles in terms of $\rho$ and $\sigma$ as
\begin{eqnarray}
\sin \alpha_t =\sqrt{\frac{1-\sigma^2}{\rho^2-\sigma^2}}\\
\sin \alpha_h =\sqrt{\frac{1-\sigma^2}{\rho^2-\sigma^2}}\rho.
\end{eqnarray}
By inserting these parameters back into the expression for the energy, we find the energy in explicit form as a function of the coupling constants:
\begin{widetext}\begin{eqnarray}
\frac{E_{4a}}{NJ_{t}} & = & -1+\frac{j_{th}^{2}}{2j_{h}}-\frac{5}{3}j_{th}^{2}-\frac{j_{th}^{4}}{4j_{h}^{2}}+\frac{j_{th}^{3}\sqrt{j_{th}^{2}-4j_{h}}}{4j_{h}^{2}}+\frac{12j_{h}^{2}+j_{th}^{2}\left(5j_{h}-12\right)}{12j_{h}+9j_{th}^{2}}\nonumber \\
 &  & +\frac{8j_{h}(2j_{h}-1)+2j_{th}^{2}(4j_{h}+1)-5j_{th}^{4}}{\sqrt{j_{th}^{2}-4j_{h}}\left(4j_{h}+3j_{th}^{2}\right)}j_{th}+\frac{2j_{th}}{\sqrt{j_{th}^{2}-4j_{h}}}\nonumber \\
 &  & -4\sqrt{2}\frac{j_{th}^{2}}{j_{th}-\sqrt{j_{th}^{2}-4j_{h}}}\sqrt{\frac{\left(j_{th}^{2}-j_{th}\sqrt{j_{th}^{2}-4j_{h}}-2j_{h}(j_{h}+1)\right)}{\left(-j_{th}^{4}+j_{th}^{3}\sqrt{j_{th}^{2}-4j_{h}}+2j_{th}^{2}j_{h}+8j_{h}^{2}\right)}}\nonumber \\
 &  & \times\sqrt{\frac{\left(3j_{th}^{4}-3j_{th}^{3}\sqrt{j_{th}^{2}-4j_{h}}+8j_{h}^{2}(1+j_{h})+4j_{th}j_{h}(j_{h}+2)\sqrt{j_{th}^{2}-4j_{h}}-2j_{th}^{2}j_{h}(j_{h}+7)\right)}{\left(-3j_{th}^{4}+8j_{th}^{2}j_{h}+16j_{h}^{2}\right)}}.
\label{energycone}
\end{eqnarray}
\end{widetext}
The necessary condition for the existence of this phase is that the coupling constants are such that all three quantities $\alpha_t,\alpha_h,\theta$ are real. As a first condition we have
\begin{eqnarray}
j_{th}^2\geq4j_h
\end{eqnarray}
in order to make the square root expression in $\rho,\sigma$ and $\theta$ real.

The requirement that $\alpha_t$ be real translates into 
\begin{eqnarray}
0\leq\frac{1-\sigma^2}{\rho^2-\sigma^2}\leq1. \label{ineqph4}
\end{eqnarray}
From the form of $\rho$ and $\sigma$ in terms of the coupling constants it is straightforward to show that $\sigma\leq\rho$. With this the inequality (\ref{ineqph4}) results in the final requirement
\begin{eqnarray}
\sigma\leq 1\leq \rho.
\end{eqnarray}
There are no further conditions, since $\alpha_h$ is guaranteed to be real if $\alpha_t$ is.

We return to the case $\rho=0$, which is also one of the solutions of Eqs.~\eqref{eq:20}-\eqref{eq:22} with
\begin{eqnarray}
\rho&=&0\rightarrow \alpha_h=0\\
\sigma&=&\frac{1}{2j_{th}}\rightarrow \cos\alpha_t=2j_{th}\\
\theta&=&\frac{2\pi}{3}.
\end{eqnarray}
The energy in terms of the coupling constants has a simpler form
\begin{eqnarray}
E_{4b}=-\frac{3}{2}NJ_t -NJ_h-2N\frac{J_{th}^2}{J_t}
\end{eqnarray}
This is the other double cone phase that we have denoted $(4b)$ in the phase diagram. The necessary condition for this phase to exist, which follows from the requirement that $\alpha_t$ must be real, is
\begin{eqnarray}
j_{th}\leq\frac{1}{2}.
\end{eqnarray}

\subsection{Incommensurate spiral phase $(5)$}
\label{sec:spiral-phase-5}
We calculate the energy with the spiral angle $\theta$ as a variational parameter. The variational expression for the energy is
\begin{align}
  \label{eq:16}
E(\theta) & =  -NJ_t+NJ_h-2NJ_{th}+2NJ_t\cos^{2}\theta \nonumber \\
 &+\left(2NJ_t+2NJ_h-4NJ_{th}\right)\cos\theta.  
\end{align}
We minimize with respect to $\theta$ and obtain
\begin{eqnarray*}
\cos\theta & = & -\frac{1}{2}\left(1+\frac{J_{h}}{J_{t}}-2\frac{J_{th}}{J_{t}}\right)\\
E_5 & = & -\frac{3}{2}NJ_t-\frac{N}{2J_t}\left(J_{h}-2J_{th}\right)^{2}
\end{eqnarray*}
as minimizing angle and energy. Since $\theta$ must be real, it follows that 
\begin{eqnarray*}
\left|1+j_h-2j_{th} \right|\leq2
\end{eqnarray*}
must hold for the existence of the spiral phase.  
\subsection{Incommensurate alternating spiral phase (XY I)}
\label{sec:altern-spir-xy}
The total variational energy is given by
\begin{align}
  \label{eq:27}
  E (\theta) &= -NJ_t + 2 NJ_{th}+NJ_h +2NJ_t (\cos^2\theta -\cos \theta) \nonumber \\
& \qquad  -4NJ_{th} \cos \theta -2 J_h \cos \theta \,.
\end{align}
The energy is minimized for 
\begin{eqnarray}
\cos \theta = \frac{1}{2} + \frac{J_{th}}{J_t}+\frac{J_h}{2J_t}
\end{eqnarray}
which yields the energy of phase XY I:
\begin{eqnarray}
E_{\text{XY I}}=-\frac{3}{2}NJ_t - N\frac{(J_h+2J_{th} )^2}{2J_t}.
\end{eqnarray}

\subsection{Canted ferromagnetic phase  (XY II)}
\label{sec:cant-ferr-xy}
This phase is related to phase (4a) and the minimization problem may be solved in close analogy with it. The energy is given in terms of the parameters $\alpha_t$ and $\alpha_h$:
\begin{align}
  \label{eq:28}
E&=NJ_t +N J_h + 2NJ_t \cos 2\alpha_t +2NJ_h \cos 2\alpha_h \nonumber \\
& - 2 NJ_{th} \left[\cos(\alpha_t+\alpha_h)+2\cos(\alpha_t-\alpha_h)\right]\,.  
\end{align}
The minimization conditions are given by
\begin{eqnarray}
2j_h \rho \sigma = j_{th} (\sigma+\rho)\\
3\sigma-\rho=\frac{4}{j_{th} }.
\end{eqnarray}
These equations are solved by 
\begin{align}
  \label{eq:29}
\sigma_{\pm} &\equiv \frac{\cos \alpha_{{h}}}{\cos \alpha_t} \\
&= \frac{2}{3j_{th}} + \frac{j_{th}}{3j_h} \pm \sqrt{\left(\frac{2}{3j_{th}} + \frac{j_{th}}{3j_h}\right)^2-\frac{1}{j_h}}\nonumber \\
\label{eq:30}
\rho &\equiv \frac{\sin \alpha_{{h}}}{\sin \alpha_t}= 3\sigma-\frac{4}{j_{th}}\,.
\end{align}
The positive sign of $\sigma$ leads to a solution with an energy that is always larger than that of the other planar phases. Hence we will disregard this configuration and only focus on the negative sign solution $\sigma_-$ (and drop the subscript on $\sigma$). As a first condition we require that the expression under the square root in the formula for $\sigma$ must not be negative. This is easily shown to be satisfied in exactly two disjoint regions of parameter space
\begin{eqnarray}
  \label{eq:31}
j_{th}^2 &\leq& j_h \\
\label{eq:32}
j_h&\leq&j_{th}^2/4 \,.
\end{eqnarray}
Furthermore, we require, as we did for phase $(4a)$, that $\alpha_t$ is real. If $\alpha_t$ is real, so will be $\alpha_h$. This condition will hold if either one of the two inequalities
\begin{align}
  \label{eq:33}
\sigma& < 1 < \rho\\
\label{eq:34}
\rho & < 1 < \sigma  \,.
\end{align}
is satisfied. From the form of $\sigma$ and $\rho$ in terms of the coupling constants we can show that the first case does not exist for any choice of coupling constants. Thus we require only the second inequality \eqref{eq:34}.

Finally, we compute the energy from the values of the angles $\alpha_t$ and $\alpha_h$, which are known in terms of $\sigma$ and $\rho$, and find
\begin{widetext}
  \begin{align}
    \label{eq:35}
E_{\text {XY II}}&=\frac{9 j_{th}^2}{4} \sqrt{\frac{  \frac{2 \left(j_{h}^2+1\right) j_{th}^4}{j_{h}^2}+8 j_{h}^2-\frac{9j_{h}^2+2j_{h}+9}{j_{h}}j_{th}^2+8+\left[\frac{2 \left(j_{h}^2-1\right) j_{th}^3 }{j_{h}}-4 \left(j_{h}^2-1\right) j_{th} \right]\sqrt{\frac{j_{th}^2-5 j_{h}}{j_{h}^2}+\frac{4}{j_{th}^2}  }}{j_{th}^4-5 j_{h} j_{th}^2+4 j_{h}^2}}\nonumber \\
&-\frac{(j_{h}-1) \left(8 j_{h}-5 j_{th}^2\right) \left(2 j_{h}+j_{th}^2\right)}{4 j_{h}\sqrt{j_{th}^4-5 j_{h} j_{th}^2+4j_{h}^2}}+\frac{5 (j_{h}+1) j_{th}^2}{4 j_{h}}+j_{h}+1 \,.    
  \end{align}
\end{widetext}


%

\end{document}